\pdfoutput=1

\documentclass[3p]{elsarticle}

\usepackage{graphicx}
\usepackage{amssymb}
\usepackage{tabularx}
\usepackage{booktabs} 
\usepackage{siunitx}
\usepackage{units}
\usepackage{placeins}
\usepackage{color}
\usepackage{soul}
\usepackage{breqn}
\usepackage{hyperref}
\usepackage{textcomp}

\hypersetup{
	colorlinks   = true, 
	urlcolor     = blue, 
	linkcolor    = blue,
	citecolor   = red
}

\biboptions{sort&compress}

\journal{Journal of the European Ceramic Society}

\begin{document}

\begin{frontmatter}

\title{Non-linear dissolution mechanisms of sodium calcium phosphate glasses as a function of pH in various aqueous media}

\author[label1]{Reece N. Oosterbeek \corref{cor1}}
\ead{rno23@cam.ac.uk}

\author[label1]{Kalliope I. Margaronis}
\author[label2]{Xiang C. Zhang}
\author[label1]{Serena M. Best}

\author[label1]{Ruth E. Cameron  \corref{cor1}}
\ead{rec11@cam.ac.uk}

\cortext[cor1]{Corresponding author}

\address[label1]{Cambridge Centre for Medical Materials, Department of Materials Science and Metallurgy, University of Cambridge, Cambridge, United Kingdom}
\address[label2]{Lucideon Ltd, Queens Road, Penkhull, Stoke-on-Trent, United Kingdom}

\begin{abstract}

Phosphate glasses for bioresorbable implants display dissolution rates that vary significantly with composition, however currently their mechanisms of dissolution are not well understood. Based on this systematic study we present new insights into these mechanisms.

Two-stage dissolution was observed, with time dependence initially parabolic and later linear, and a two-stage model was developed to describe this behaviour. Dissolution was accelerated by lower Ca concentration in the glass, and lower pH in the dissolution medium.

A new dissolution mechanism is proposed, involving an initial stage where diffusion-controlled formation of a conversion layer occurs. Once the conversion layer is stabilised, layer dissolution reactions become rate-limiting. Under this mechanism the transition time is sensitive to the nature of the conversion layer and solution conditions.

These results reveal the dependence of P\textsubscript{2}O\textsubscript{5}-CaO-Na\textsubscript{2}O glass dissolution on solution pH, and provide new insight into the dissolution mechanisms, particularly regarding the transition between the two dissolution stages.

\end{abstract}


\end{frontmatter}

%%%%%%%%%%%%%%%%%%%%%%%%%%%%%%%%%%%%%%%%%%%%%%%%%%%%%%%%%%%%%%%%%%%%%%%%%%%%%%%%%%%%%%%%%%%%%%%%%%%%%%%

\section{Introduction}

Phosphate-based glasses are attractive materials for medical implants, due in part to their cytocompatibility and demonstrated potential for use in soft tissue applications such as ligament and muscle scaffolds, wound healing, and promoting angiogenesis \cite{Kargozar2018, Bitar2004, Boccaccini2016, Shah2005, AbouNeel2005}. However perhaps their greatest advantage is their solubility in aqueous solutions, and the ability to tune this solubility over many orders of magnitude by tailoring the glass composition, to match the degradation lifetime of the material with tissue repair \cite{Parsons2006}. Phosphate glasses have potential applications including hard tissue engineering, as well as controlled release and antimicrobial materials \cite{Knowles2003, AbouNeel2009}, and can also be combined with polymers in fully degradable polymer-glass composites for hard or soft tissue implant materials \cite{Sharmin2016, Felfel2013, ShahMohammadi2011, Bitar2004, Gough2002, Gough2003}. The biological response of these materials is often closely linked to the dissolution rate of the glass - in order to support cell adhesion and survival, the surface must dissolve slowly enough to allow physical bonding \cite{Bitar2004, Gough2002, Gough2003, Parsons2004a}. It is clear that in any application, the degradation timescale of the material is of critical importance, therefore intimate knowledge of the dissolution behaviour of the phosphate glass is crucial.\\

Due to their similarity in composition to natural bone, calcium containing phosphate glasses are of particular interest for medical applications. Early work by Bunker et al. and Uo et al. \cite{Bunker1984, Uo1998} found that Ca ions introduced within the glass form ionic crosslinks between non-bridging oxygens of two different glass chains \cite{VanWazer1950}, enhancing the chemical durability. Later work has shown Mg and Fe to have a similar, even stronger effect \cite{Parsons2006, Franks2000, Salih2000}, with crosslinking of non-bridging oxygens by di- and tri-valent metal ions having a dominant effect on the physical properties, including dissolution rate, density, and $T_{g}$. The pH dependence of Ca-P glass dissolution has not been extensively studied, however results in similar systems such as Zn-P and Na-P glasses indicate that the rate of dissolution increases in acidic or basic solutions \cite{Bunker1984, Massera2013, Delahaye1998}, with acidic solutions accelerating dissolution by disrupting ionic crosslinks between phosphate chains.\\

Despite having been the subject of research for several decades, uncertainty remains over the mechanisms occurring during phosphate glass dissolution. Several studies \cite{Bunker1984, Ma2018, Knowles2003} have observed an initial dissolution stage with parabolic time dependence, in P\textsubscript{2}O\textsubscript{5}-CaO-Na\textsubscript{2}O glasses, which Bunker et al. attributed to water diffusion and formation of a surface hydration layer \cite{Bunker1984}. This initial stage has also been observed in P\textsubscript{2}O\textsubscript{5}-FeO-Fe\textsubscript{2}O\textsubscript{3}-Na\textsubscript{2}O glasses \cite{Ma2017}, however other works on P\textsubscript{2}O\textsubscript{5}-CaO-Na\textsubscript{2}O glasses have not observed this initial non-linear stage \cite{Delahaye1998, Franks2000}, suggesting that it may be related to the changing ionic strength of the solution rather than diffusion and ion exchange \cite{Sharmin2017}. The later stage of dissolution, with linear time dependence, is considered to be controlled by the reaction of the hydrated layer at the glass-solution interface \cite{Ma2017, Ma2018, Sharmin2017}. The cause of the transition between these stages is not yet well understood, but the transition time has been observed to be roughly correlated to the durability of the glass \cite{Bunker1984} (i.e. the overall resistance to dissolution).\\ 

In this work we investigate the dissolution behaviour of a set of P\textsubscript{2}O\textsubscript{5}-CaO-Na\textsubscript{2}O glasses in deionised (DI) water and phosphate-buffered saline (PBS). In order to gain insight into the mechanisms of dissolution, we apply a two-stage model similar to that used by Ma et al. \cite{Ma2017, Ma2018} to describe the mass loss of Na-Fe-P and Na-Ca-P glasses, based on an initial parabolic time dependence, followed by later linear dissolution. To the authors' knowledge, this is the first work to quantitatively compare the non-linear mechanisms of P\textsubscript{2}O\textsubscript{5}-CaO-Na\textsubscript{2}O glass dissolution in DI water and higher ionic strength PBS, across a range of pH values, and provides new insight into the two stages of dissolution, and the cause of the transition between them. The effect of pH is especially important when considering phosphate glasses as a component in polymer-glass composites, due to the acidification that can result from degradation of commonly used degradable polymers. To simulate the conditions phosphate glass may experience in a polymer-glass composite, we conducted dissolution experiments in PBS with added lactic acid (the acidic degradation product of common degradable polymers) to alter the solution pH, and determine the effect on the dissolution rate and mechanisms. Previous works have speculated that the transition between the two dissolution stages may be related to the nature of the conversion layer (also referred to as an alteration layer \cite{Ma2017}), here we discuss how the formation of different conversion layer species across a range of solution conditions can affect the transition behaviour.

%%%%%%%%%%%%%%%%%%%%%%%%%%%%%%%%%%%%%%%%%%%%%%%%%%%%%%%%%%%%%%%%%%%%%%%%%%%%%%%%%%%%%%%%%%%%%%%%%%%%%%%

\section{Materials and Methods}

\subsection{Glass preparation}
Glasses with nominal compositions of (P\textsubscript{2}O\textsubscript{5})\textsubscript{90-$x$}(CaO)\textsubscript{$x$}(Na\textsubscript{2}O)\textsubscript{10}, where $x$ = 40, 45, 50, were prepared using Na\textsubscript{2}CO\textsubscript{3}, CaCO\textsubscript{3}, and NH\textsubscript{4}H\textsubscript{2}PO\textsubscript{4}, referred to subsequently by codes P50Ca40, P45Ca45, and P40Ca50 for $x$ = 40, 45, 50 respectively. The precursors were melted and degassed in a kiln (SBSC-1500L, Kilns and Furnaces Ltd., Stoke-on-Trent, UK) at 1230 - 1360\si{\degree}C using a vitreous silica crucible, and then quenched quickly to room temperature. Glass discs (10 $\times$ 2 mm) were cast by pouring re-melted glass (at 1100 - 1230\si{\degree}C) into a graphite mould preheated to 400\si{\degree}C. Cast discs were then annealed at 400\si{\degree}C and left to cool slowly to room temperature, to remove any residual stresses.

\subsection{Characterisation}
Glass density was measured using an AccuPyc 1330 gas pycnometer (Micromeritics Instrument Corporation, USA). DSC (Differential scanning calorimetry) was carried out with a TA Instruments SDT Q600, using 10 - 25 mg of glass powder in an alumina pan, and heating from 25 to 1000\si{\degree}C at 20\si{\degree}C/min. Thermal properties were analysed using the TA Universal Analysis software. The data were baseline corrected by subtracting the response of an empty alumina pan, and glass transition temperatures were taken at the inflection point. XRD (X-ray diffraction) was carried out using a Bruker D8 Advance diffractometer with Cu K-shell radiation. For confirming the amorphous nature of the disks, a 2$\theta$ range of 10-60\textdegree with a 0.02\textdegree step size and dwell time of 0.5 s/step was used, while a 2$\theta$ range of 20-50\textdegree with a 0.1\textdegree step size and dwell time of 10 s/step was used to maximise the signal/noise ratio for partially dissolved samples. SEM (Scanning electron microscopy) and EDX (Energy dispersive X-ray spectroscopy) were performed using a CamScan MX2600 FEG-SEM and Phenom ProX SEM, using an accelerating voltage of 10 kV for SEM and 15 kV for EDX.
Prior to SEM/EDX, samples were sputter coated with approximately 10 nm of gold/palladium to prevent charging, using an Emitech K550 sputter coater (40 mA deposition current for 2 minutes, under an Argon atmosphere). The Q\textsuperscript{n} distribution in the glass network was measured by \textsuperscript{31}P solid-state magic angle spinning (MAS) NMR, using a Bruker Avance 300 spectrometer at 121 MHz and a 4 mm MAS probe. 1D spectra were fitted using TopSpin 4.0.5 (Bruker Ltd.) to quantify the Q\textsuperscript{n} species present. 

\subsection{Dissolution testing}
Dissolution tests were carried out in DI water (Type I, 18.2 M$\Omega$) and PBS (phosphate-buffered saline, pH = 7) (Gibco, Thermo Fisher Scientific Inc., USA), as well as PBS adjusted to pH = 3, and pH = 5 by adding lactic acid ($\geq$85\%, Sigma-Aldrich 252476). Glass discs were immersed in 15 mL of the chosen solution (SA/V ratio 15 m\textsuperscript{-1}, chosen for comparison with published literature) in conical-bottomed vials (to ensure all surfaces were exposed to solution), and incubated at 37\si{\degree}C without agitation. Solution pH was measured using a Hanna HI 4222 pH meter with HI 1131B pH electrode and three point calibration. Ca\textsuperscript{2+} ion concentration was measured using an ISE (ion selective electrode - Sentek 361-75 Calcium Combination ISE), calibrated using a modified Nernst equation \cite{Midgley1977} with standard solutions diluted with DI water from a 0.1 mol L\textsuperscript{-1} calcium standard (Hanna HI 4004-01). To measure glass disc mass during dissolution, discs were removed from solution, rinsed with DI water, dabbed dry, and weighed (using a Sartorius BP61 balance with internal calibration, d = 0.1 mg), before being returned to the solution for further dissolution. Digital photographs were also taken to record visible changes in morphology. Before SEM imaging, partially dissolved discs were removed from solution, rinsed with DI water, dabbed dry, and stored in a desiccator to dry over several days.

%%%%%%%%%%%%%%%%%%%%%%%%%%%%%%%%%%%%%%%%%%%%%%%%%%%%%%%%%%%%%%%%%%%%%%%%%%%%%%%%%%%%%%%%%%%%%%%%%%%%%%%

\section{Calculation}

\begin{figure}[h]
	\centering
	\includegraphics[width=0.5\linewidth]{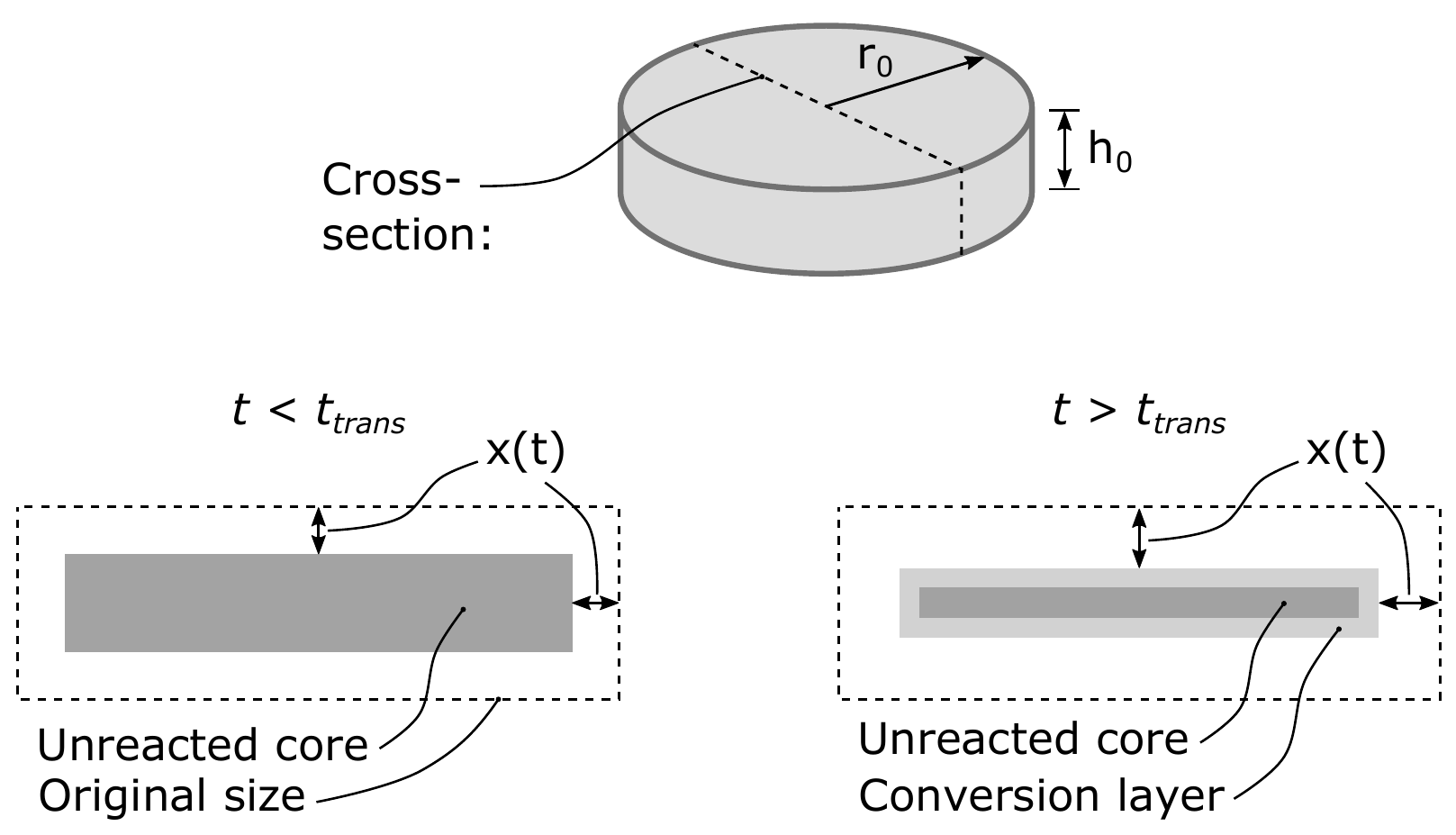}	
	\caption{Schematic showing the model geometry for glass dissolution, and two-stage dissolution model as described by Eqs. \ref{eq:Mod-x} and \ref{eq:Mod-a}.}
	\label{fig:MechSchem}
\end{figure}

A two-stage dissolution model is used to describe the glass mass loss in a similar fashion to the work of Ma et al. \cite{Ma2017}, however here we adapt this model for use in a disc geometry  by considering the thickness of a dissolved layer $x$. In the initial diffusion controlled stage (before the transition time $t_{trans}$ is reached, i.e. $t < t_{trans}$), dissolution occurs at the surface of the disc and the interface moves towards the centre, resulting in the unreacted core shrinking and decreasing the surface and hence reaction area over time as shown in Fig. \ref{fig:MechSchem}. In this stage a 3D diffusion model (DM) with parabolic time dependence applies \cite{Rahaman2003, Gu2011}, such that:

\begin{equation}
\label{eq:Mod-kDM}
x(t < t_{trans}) = k_{DM}t^{\nicefrac{1}{2}}	
\end{equation}

In later stages ($t > t_{trans}$) the surface reaction determines the dissolution rate, with conversion reactions progressing into the disc by diffusion (as shown in Fig. \ref{fig:MechSchem}), while dissolution of the conversion layer occurs at the surface. Assuming linear reaction kinetics the dissolution progress can be described by a contracting volume model \cite{Rahaman2003, Jung2009} (CVM):

\begin{equation}
\label{eq:Mod-kCVM}
x(t > t_{trans}) = k_{CVM}t
\end{equation}

Therefore the extent of the dissolved layer over the whole course of dissolution can be described by:

\begin{equation}
\label{eq:Mod-x}
x(t) = H(t_{trans} - t)k_{DM}t^{\nicefrac{1}{2}} + H(t - t_{trans})(k_{DM}t_{trans}^{\nicefrac{1}{2}} + k_{CVM}(t - t_{trans}))
\end{equation}

where $H$ is the Heaviside step function (half-maximum convention). Applying the disc geometry, the mass fraction dissolved ($\alpha$) can be given by:

\begin{equation}
\label{eq:Mod-a}
\alpha(t) = 1 - \frac{(r_{0} - x(t))^{2}(h_{0} - x(t))}{r_{0}^{2}h_{0}}
\end{equation}

where $r_{0}$ and $h_{0}$ are, respectively, the initial radius and height of the disc, as shown in Fig. \ref{fig:MechSchem}. Equations 3 and 4 were used to model the mass loss of phosphate glasses, and were fitted to experimental data using non-linear least squares regression in Matlab.

%%%%%%%%%%%%%%%%%%%%%%%%%%%%%%%%%%%%%%%%%%%%%%%%%%%%%%%%%%%%%%%%%%%%%%%%%%%%%%%%%%%%%%%%%%%%%%%%%%%%%%%

\FloatBarrier
\section{Results}

\subsection{Characterisation}

A summary of the properties of the prepared glass discs is shown in Table \ref{tab:Char}. The EDX-derived compositions indicated that the P\textsubscript{2}O\textsubscript{5} and CaO content was consistent with that expected from the feed, however the Na\textsubscript{2}O was lower than the 10 mol\% expected, and the presence of SiO\textsubscript{2} (which was not included in the feed) was detected. O/P ratios are also given for comparison with other works. From \textsuperscript{31}P MAS-NMR Q\textsuperscript{1} and Q\textsuperscript{2} phosphate tetrahedra (1 and 2 bridging oxygens respectively) were seen, at positions from -8.2 to -8.5 ppm for Q\textsuperscript{1}, and from -23.2 to -25.7 ppm for Q\textsuperscript{2} \cite{Brow2000, Fletcher1993, Gras2016}. As the Ca content increased the Q\textsuperscript{1}/Q\textsuperscript{2} ratio increased, indicating depolymerisation of the network and replacement of chain forming Q\textsuperscript{2} groups with chain terminating Q\textsuperscript{1} groups. XRD did not detect the presence of any diffraction peaks, indicating that the glass discs used were completely amorphous.\\

\begin{table*}
	\centering
	\caption{\label{tab:Char}{Measured physical and chemical properties of the glasses produced and used in this study. Compositions are measured by EDX, and the error shown is the standard deviation from measurements of three discs from the same batch.}}
	{\renewcommand{\arraystretch}{1.2}
		{\small
			\begin{tabularx}{\linewidth}{ l X X X X X X X X}
				\toprule
				Glass code & P\textsubscript{2}O\textsubscript{5} (mol\%) & CaO (mol\%) & Na\textsubscript{2}O (mol\%) & SiO\textsubscript{2} (mol\%) & O/P ratio & $T_{g}$ (\si{\degree}C) & Q\textsuperscript{1}/Q\textsuperscript{2} & Density (g/cm\textsuperscript{3}) \\
				\midrule
				P50Ca40 & 49.8 ($\pm0.3$) & 40.1 ($\pm0.3$) & 6.4 ($\pm0.2$) & 3.7 ($\pm0.2$) & 3.041 ($\pm0.007$) & 451.6 ($\pm0.7$) & 0.027 ($\pm0.003$) & 2.6147 ($\pm0.0009$) \\
				P45Ca45 & 44.8 ($\pm0.3$) & 45.5 ($\pm0.4$) & 6.9 ($\pm0.2$) & 2.8 ($\pm0.2$) & 3.147 ($\pm0.009$) & 466.9 ($\pm0.7$) & 0.36 ($\pm0.02$) & 2.7062 ($\pm0.0003$) \\
				P40Ca50 & 40.4 ($\pm0.2$) & 50.8 ($\pm0.4$) & 6.4 ($\pm0.2$) & 2.4 ($\pm0.2$) & 3.267 ($\pm0.007$) & 489.5 ($\pm0.5$) & 0.93 ($\pm0.01$) & 2.784 ($\pm0.001$) \\
				\bottomrule 
			\end{tabularx}
	}}
\end{table*}

\subsection{Dissolution in DI water}

\begin{figure}[h]
	\centering
	\includegraphics[width=0.5\linewidth]{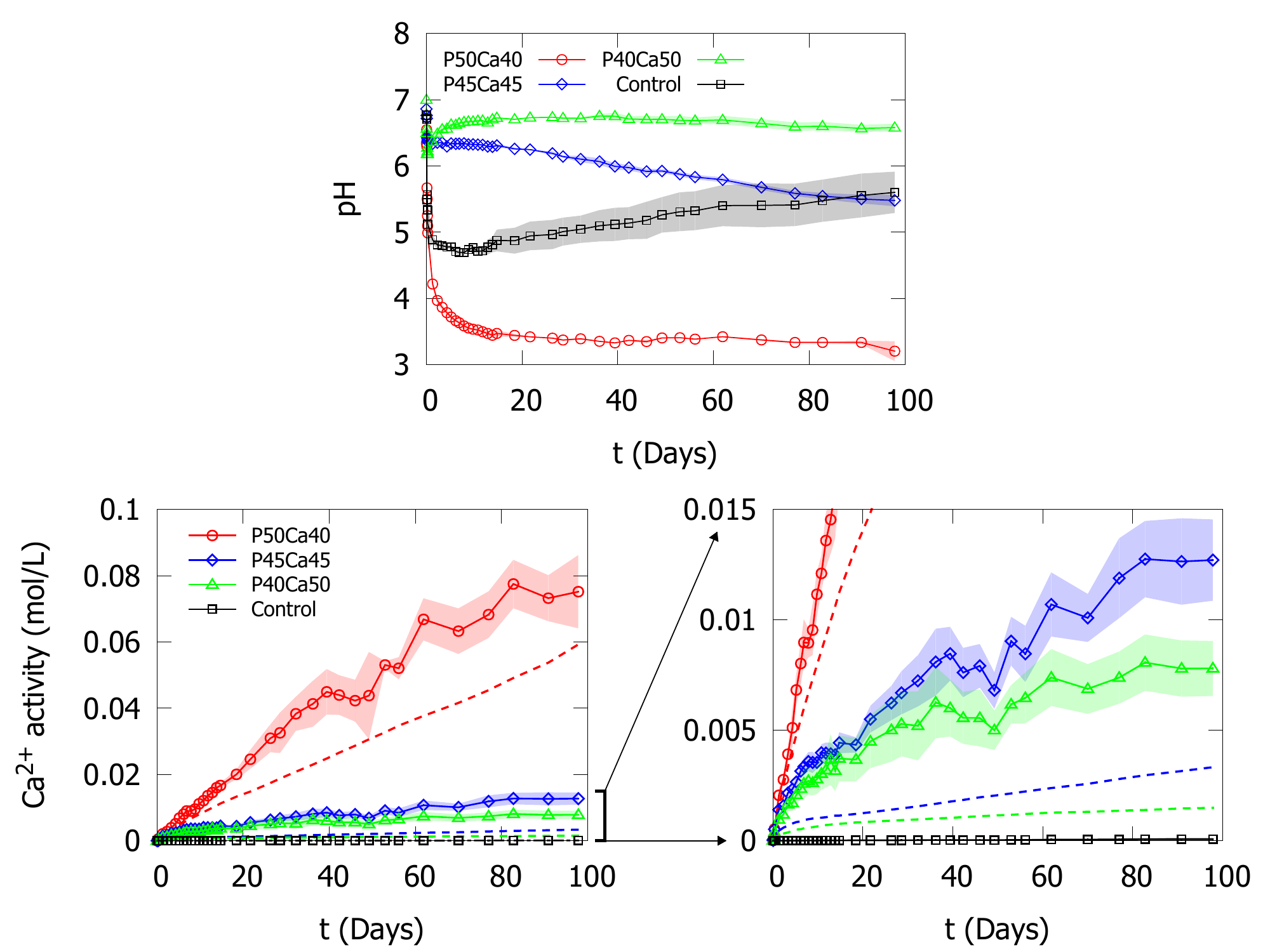}	
	\caption{Solution pH (top) and Ca\textsuperscript{2+} activity (bottom) for phosphate glasses dissolving in DI water at 37\si{\degree}C. Solid lines indicate Ca\textsuperscript{2+} activity measured using an ISE, while dotted lines indicate Ca\textsuperscript{2+} activity calculated based on measured mass loss assuming congruent dissolution. Shaded region denotes standard deviation for n = 3 measurements.}
	\label{fig:DI_pHaCa}
\end{figure}

Solution pH and Ca\textsuperscript{2+} ion activity changes as the glasses dissolve in DI water are seen in Fig. \ref{fig:DI_pHaCa}. The control solution (DI water only) suffered a rapid decrease in pH to around 4.7 due to dissolution of atmospheric CO\textsubscript{2}, followed by a gradual increase. This gradual increase can be attributed to contamination of the solution with residual ions from the electrode - in spite of thorough washing between samples, some transfer was unavoidable, and resulted in a pH increase for the control solution which has very low ionic concentration and therefore is very sensitive to changes.\\

\begin{figure}[h!]
	\centering
	\includegraphics[width=0.5\linewidth]{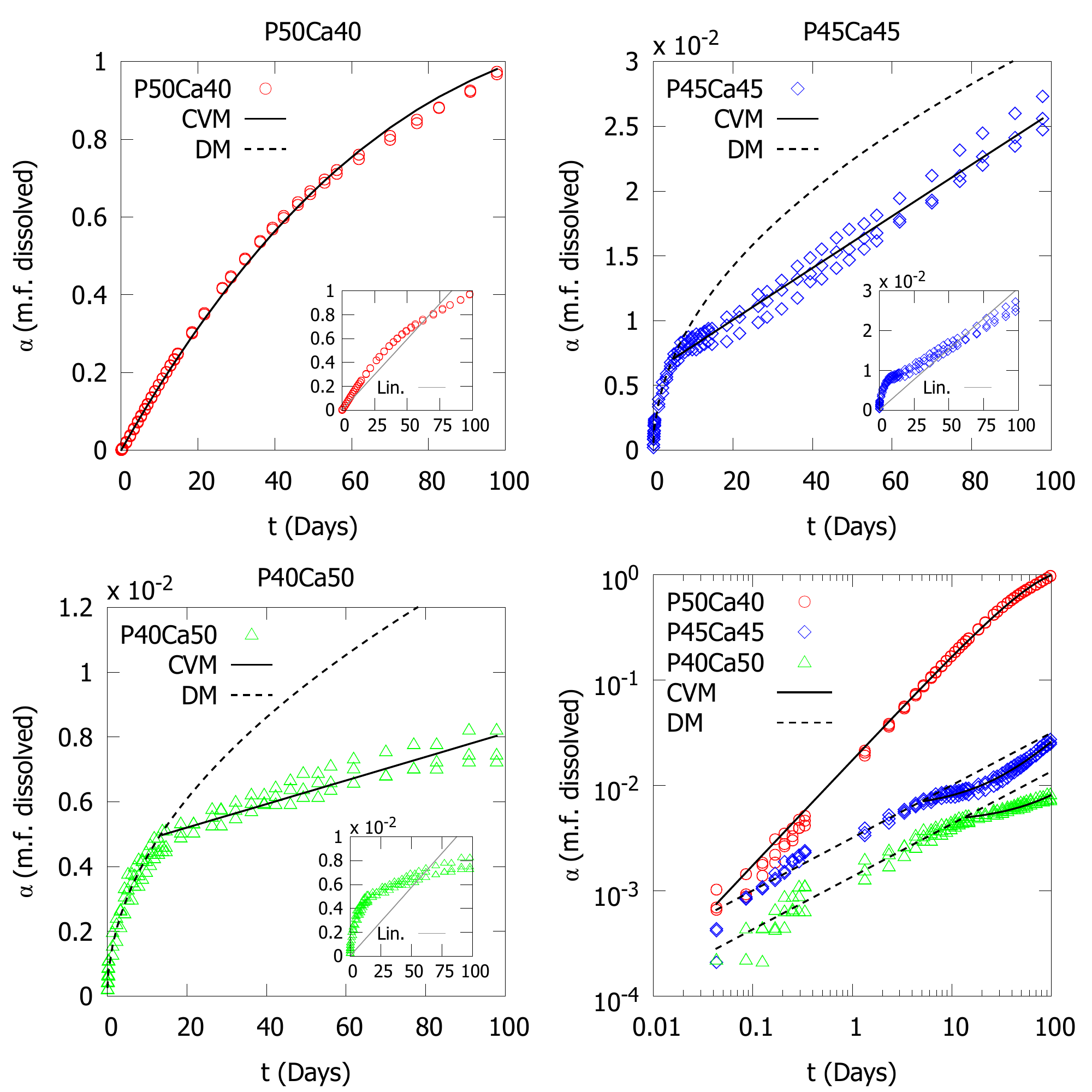}
	\caption{Mass fraction dissolved during dissolution in DI water at 37\si{\degree}C for phosphate glasses, fitted to two-stage model with diffusion stage (dotted line - DM) and contracting volume stage (solid line - CVM). Inset plots show the same data fitted to a linear dissolution rate. Bottom right shows the same data combined onto a log-log scale for comparison.}
	\label{fig:DI_mass}
\end{figure}

Changes in measured Ca\textsuperscript{2+} ion activity indicate that the level of Ca released from the glass (Fig. \ref{fig:DI_pHaCa}) was the inverse of Ca content, with the lowest Ca glass (P50Ca40) releasing Ca the fastest - clearly Ca release is more closely linked to the dissolution behaviour (see below) than actual Ca content. Dotted lines in this graph indicate the Ca\textsuperscript{2+} activity calculated from the mass loss data (using the Truesdell-Jones model \cite{Truesdell1974, Garrels1965, Drever1982}) in Fig. \ref{fig:DI_mass}, assuming congruent dissolution. In all cases the measured Ca release was significantly higher than that expected from the mass loss, indicating Ca was leached from the bulk glass.\\

The mass loss of phosphate glasses during dissolution in DI water (Fig. \ref{fig:DI_mass}) shows that glasses with higher Ca content dissolved more slowly. It can also be seen that linear dissolution models (inset plots) did not describe this data adequately. By contrast the two-stage model, incorporating a diffusion stage (DM) and contracting volume stage (CVM), fitted the data well, with parameters given in Table \ref{tab:glassdiss} and Fig. \ref{fig:Mass_fits}. Rate constants ($k_{DM}, k_{CVM}$) decreased with increasing Ca, consistent with slower dissolution, while the transition time $t_{trans}$ increased. The initial diffusion limited stage was not observed for the P50Ca40 glass, therefore $t_{trans}$ can only be given as lower than the first non-zero timepoint.\\

\begin{figure}[htb]
	\centering
	\includegraphics[width=0.75\linewidth]{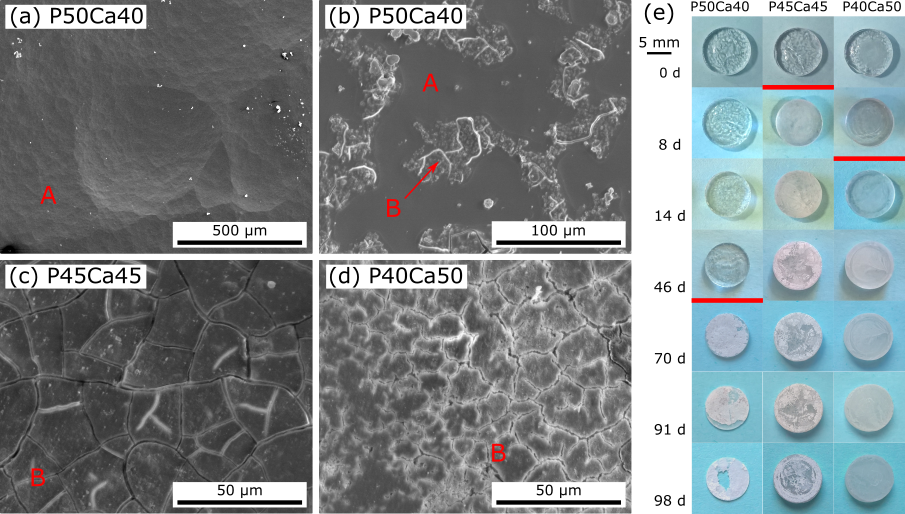}
	\caption{Morphology of glass discs during and after dissolution, showing SEM images after 98 days in DI water at 37\si{\degree}C for P50Ca40 (a, b), P45Ca45 (c), P40Ca50 (d), and digital photographs of glass discs at various dissolution timepoints (e). Red lines indicate the time taken for observation of an opaque layer.}
	\label{fig:DI_diss_morph}
\end{figure}

The morphology of the phosphate glasses after dissolution in DI water is shown in Fig. \ref{fig:DI_diss_morph}.  After 98 days a layer (denoted as type B) was observed on the surface of the P45Ca45 and P40Ca50 glasses. The morphology of this layer is difficult to interpret, as the cracks seen may be caused by the drying process, however the composition of this layer is of principal interest. EDX revealed this to be dominated by Ca and P, with a P:Ca:Na:Si ratio of 100:71:2:0 measured, regardless of the actual glass composition. A similar layer (B) with comparable composition was also seen partially covering the surface of the P50Ca40 glass, with the remaining area (denoted as type A) displaying etch pits. Areas where etch pits were found (only on P50Ca40) had a similar composition to the starting glass, however were depleted in Ca, with P:Ca:Na:Si ratio of 100:33:15:4, compared with a ratio of 100:40:13:4 before dissolution. This layer was also seen in macro-scale images of the glass discs, where it turned the transparent glass opaque (Fig. \ref{fig:DI_diss_morph}). The time at which this layer appeared roughly corresponded to the transition time between the two stages of mass loss (Table \ref{tab:glassdiss}) for the medium and high Ca glass (P45Ca45 and P40Ca50), however for the low Ca glass P50Ca40 transition occurred very early on, while the layer was only visible well into the dissolution process. The appearance of the glass discs shown in Fig. \ref{fig:DI_diss_morph}e is consistent with the assumption of homogeneous size reduction inherent in this model. When dissolution is nearly complete, this assumption would predict a very thin disc with slightly reduced diameter, which is consistent with the appearance of the P50Ca40 disc after 98 days in DI water (mass fraction dissolved $\alpha$ = 0.97). The hole that has appeared in the disc is a result of slight variations in disc thickness across the surface (on the scale of tens of microns) leading to complete dissolution in some places but not others.\\

\begin{figure}[h!]
	\centering
	\includegraphics[width=0.5\linewidth]{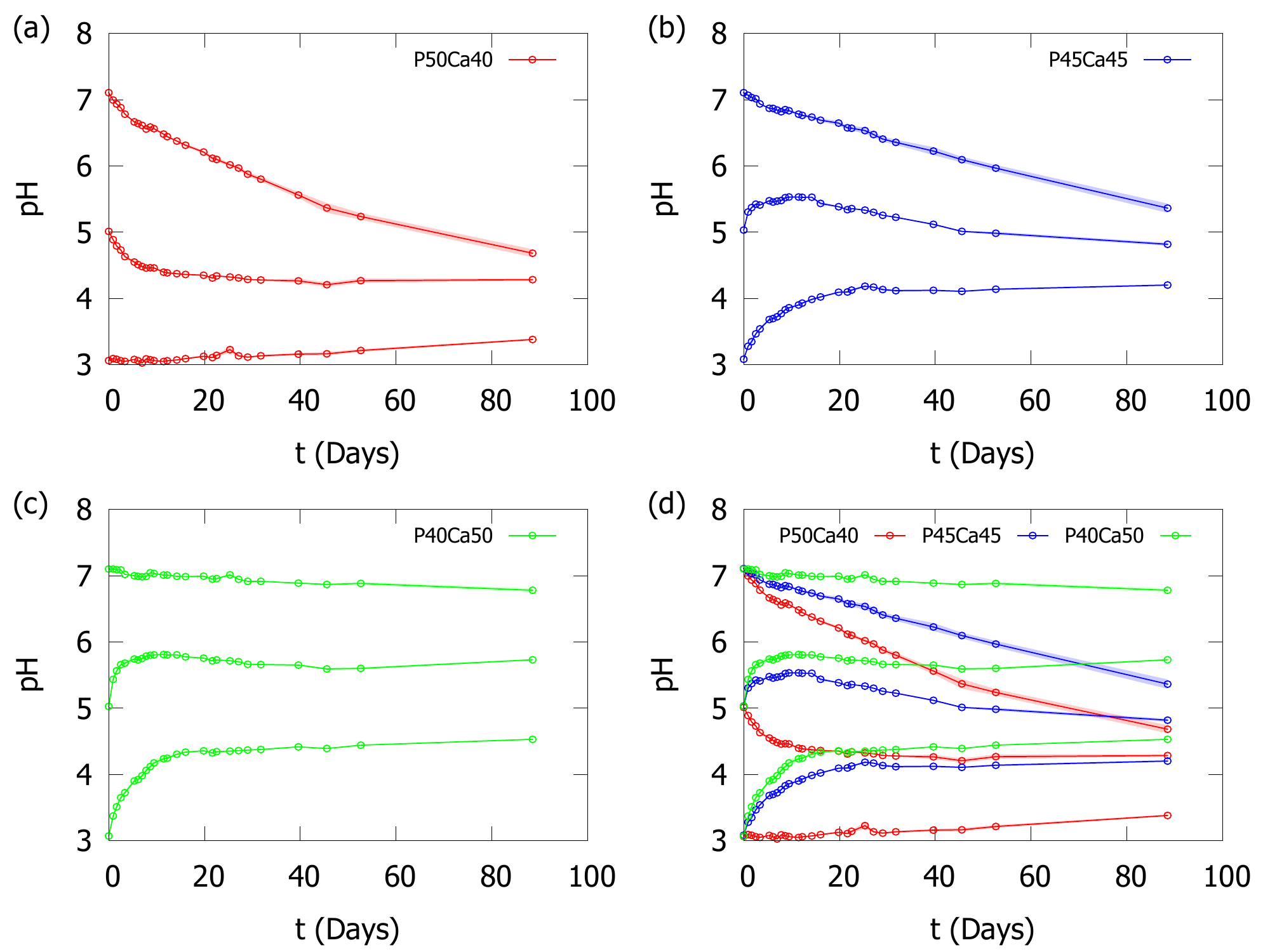}	
	\caption{Solution pH for phosphate glasses dissolving in PBS, or lactic acid-adjusted PBS at pH 3 or 5, at 37\si{\degree}C, showing P50Ca40 (a), P45Ca45 (b), P40Ca50 (c), and all three glasses together for comparison (d). Shaded region denotes standard deviation for n = 3 measurements.}
	\label{fig:PBS_pH}
\end{figure}

\begin{figure}[h!]
	\centering
	\includegraphics[width=0.5\linewidth]{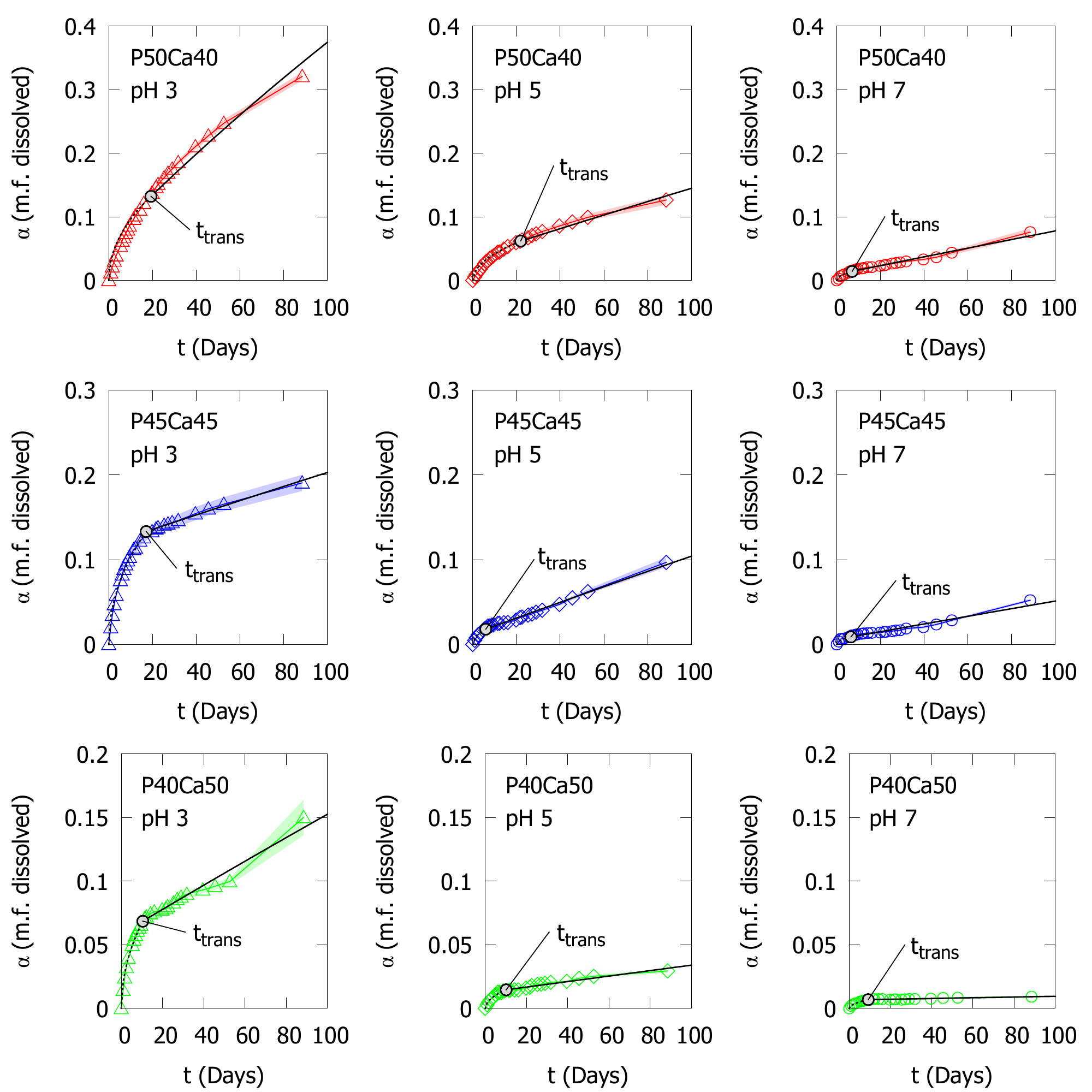}	
	\caption{Mass fraction dissolved during dissolution in PBS, or lactic acid-adjusted PBS at pH 3 or 5, at 37\si{\degree}C. Shaded region denotes standard deviation for n = 3 measurements. Black line shows fitted model according to Eq. \ref{eq:Mod-a}, with dotted and solid regions denoting the diffusion and contracting volume stages respectively.}
	\label{fig:PBS_mass}
\end{figure}

\subsection{Dissolution in PBS}

The evolution of solution pH during phosphate glass dissolution in PBS and pH-adjusted PBS is shown in Fig. \ref{fig:PBS_pH}. The trend for different glass compositions is consistent with dissolution in DI water (Fig. \ref{fig:DI_pHaCa}), with higher Ca glasses leading to higher solution pH. In general, regardless of initial pH, the dissolution of the glass in PBS appeared to result in a slow convergence of the solution pH towards a similar value for each glass composition as the glass dissolution slowly dominated over the initial solution conditions. The pH of the empty solution (PBS) was measured to be stable at pH $\approx 7$ over the experimental timescale used.\\

The mass loss of phosphate glasses in PBS and pH-adjusted PBS is shown in Fig. \ref{fig:PBS_mass}, with fitted curves from the two-stage model in Table \ref{tab:glassdiss}. Again it is clear that glasses with higher Ca content dissolved more slowly, in accordance with earlier results in DI water. The effect of pH on the general mass loss trend is also clear; reduced pH accelerated mass loss for all glass compositions.\\

\begin{figure}[htb]
	\centering
	\includegraphics[width=0.5\linewidth]{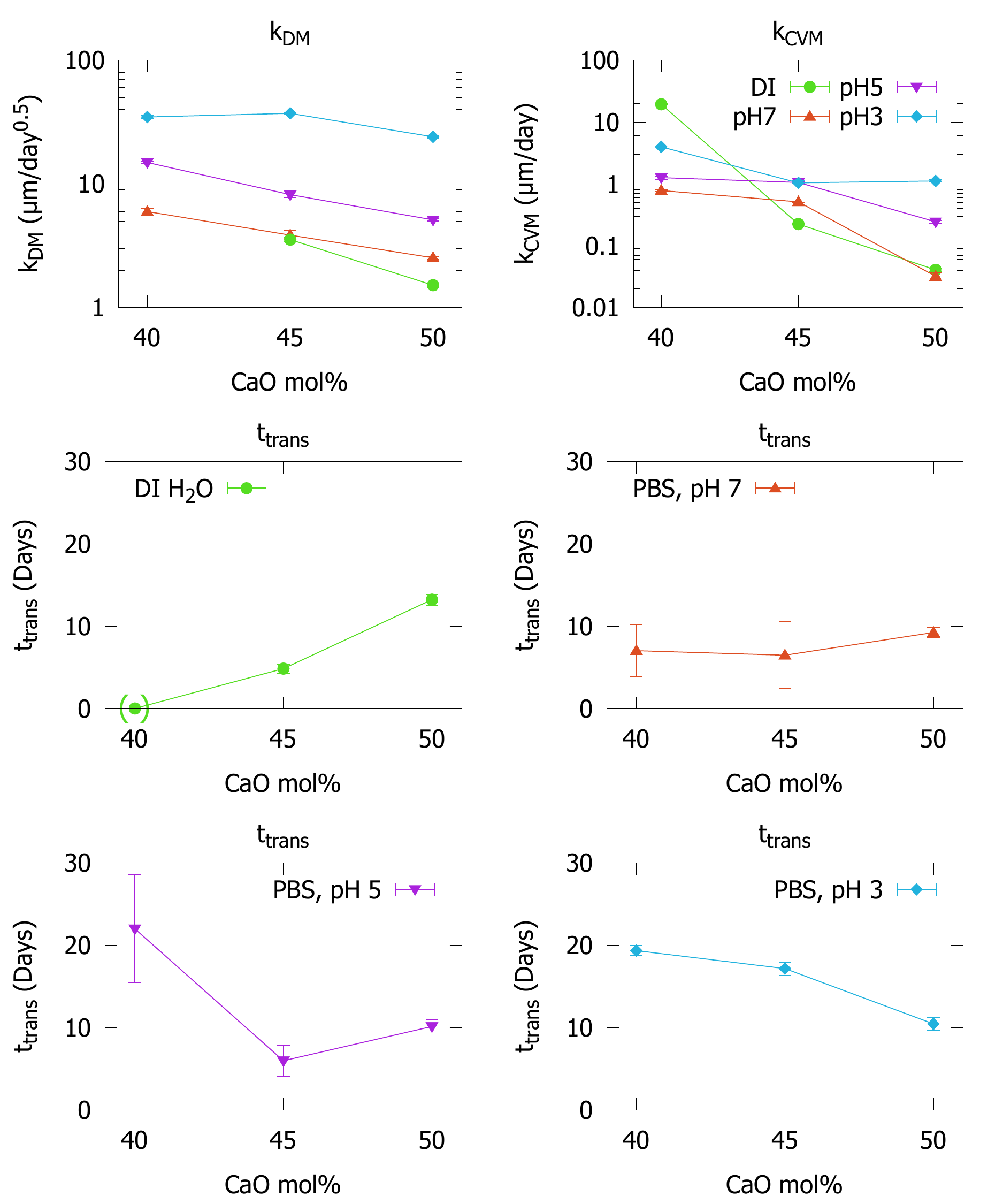}	
	\caption{Fitted parameters ($k_{DM},\ k_{CVM},\ t_{trans}$) for the two-stage model for phosphate glass dissolution in DI water, PBS, or lactic acid-adjusted PBS at pH 3 or 5, at 37\si{\degree}C C. Error bars denote standard deviation for n = 3 measurements, brackets () denote transition occurring immediately, with no diffusion-limited stage observed.}
	\label{fig:Mass_fits}
\end{figure}

The fitted parameters for the two-stage model (Table \ref{tab:glassdiss} and Fig. \ref{fig:Mass_fits}) showed significant variation across glass composition and solution conditions. The rate constant for the diffusion controlled stage ($k_{DM}$) showed a clear decreasing trend for increasing Ca content, and also increased significantly as the solution pH decreased. Similar trends, although less clear, were also seen for the rate constant for the reaction controlled stage ($k_{CVM}$). Variation in the transition time ($t_{trans}$) between the two stages was also seen across different glass compositions and solution conditions. In DI water, $t_{trans}$ increased with increasing Ca content, however this trend was absent in PBS with a similar $t_{trans}$ observed for all glass compositions, and reversed in pH-adjusted PBS, with a decreasing $t_{trans}$ with increasing Ca content.\\

\begin{figure}[htb]
	\centering
	\includegraphics[width=0.5\linewidth]{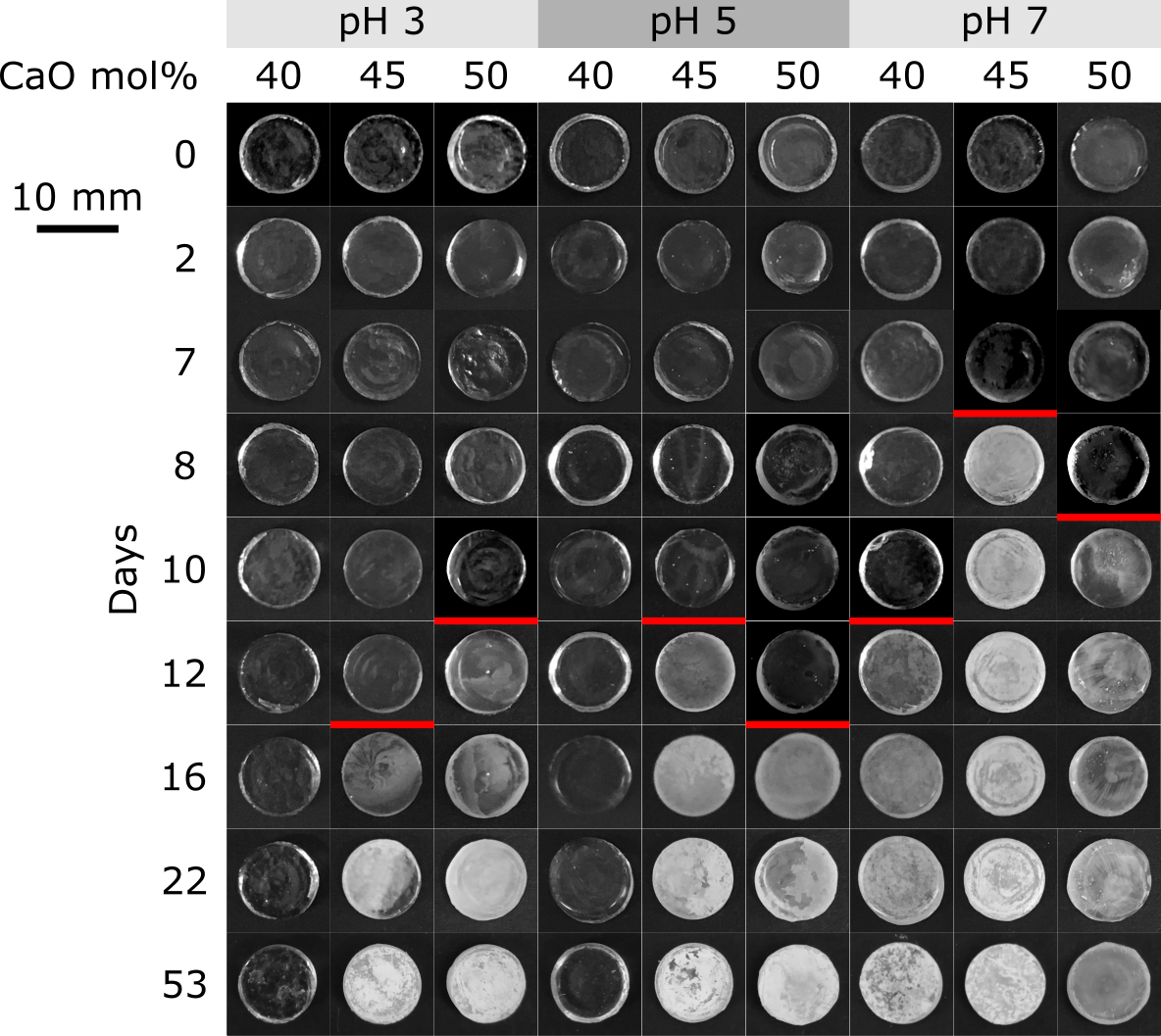}	
	\caption{Morphology of glass discs during and after dissolution in PBS, or lactic acid-adjusted PBS at pH 3 or 5, at 37\si{\degree}C. Red lines indicate the time taken for observation of an opaque layer.}
	\label{fig:PBS_morph}
\end{figure}

\begin{figure}[h!]
	\centering
	\includegraphics[width=0.5\linewidth]{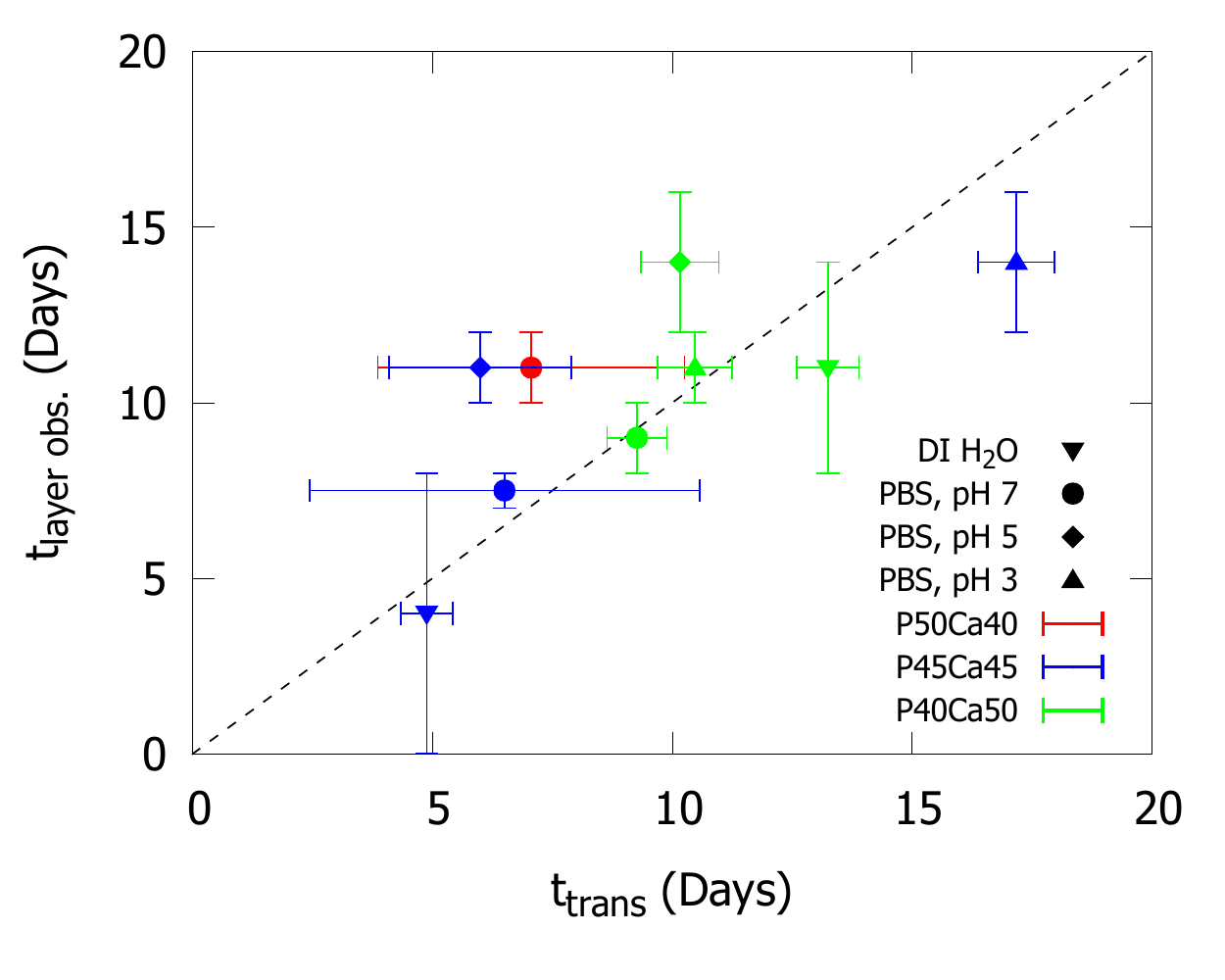}	
	\caption{Comparison of fitted $t_{trans}$ parameter, and the time taken to observe formation of an opaque layer on the glass surface. Colours denote glass compositions, and symbols denote the dissolution media, as described in the key. Dotted line represents $t_{trans} = t_{layer\ obs}$.}
	\label{fig:t_trans_comp}
\end{figure}

\begin{figure}[htb]
	\centering
	\includegraphics[width=0.5\linewidth]{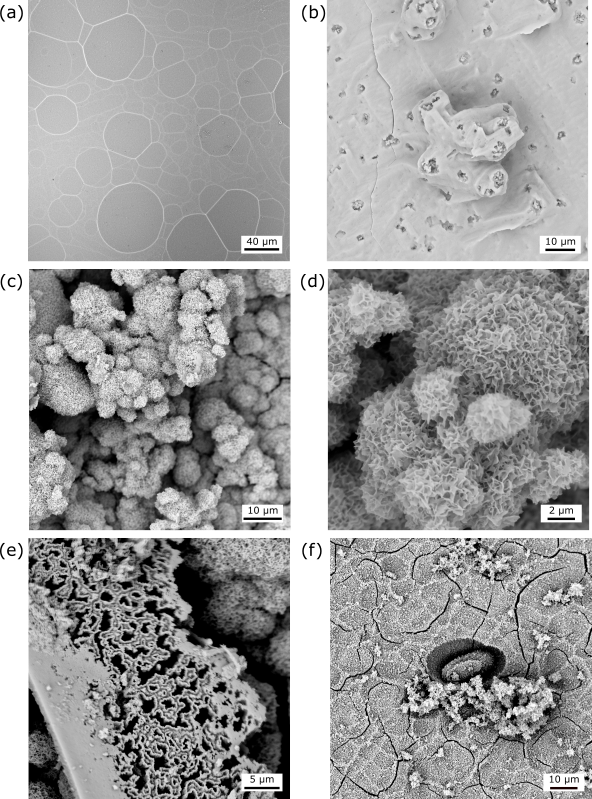}	
	\caption{SEM images of phosphate glasses during dissolution showing (a) type A behaviour (etch pits) on P45Ca45 in pH 3 below $t_{trans}$, and (b) type C behaviour (NaCl layer formation) on P40Ca50 in pH 7 above $t_{trans}$. (c-f) show type B behaviour (CaP layer formation): on P45Ca45 in pH 3 after 88 days (c, d), on P40Ca50 in pH 3 after 88 days (e), and on P40Ca50 in pH 7 above $t_{trans}$ (f).}
	\label{fig:PBS_SEM}
\end{figure}

The macro-scale morphology of glass discs during dissolution is shown in Fig. \ref{fig:PBS_morph}, where the formation of an opaque layer was observed for all glasses except for P50Ca40 in pH 3 and pH 5. The time taken for observation of this layer is also shown plotted against the fitted $t_{trans}$ parameter in Fig. \ref{fig:t_trans_comp}. Data points for P50Ca40 in DI water and PBS at pH 3 and 5 were excluded from this plot as transition occurred immediately in DI water, and no layer was observed in PBS at pH 3 and 5 within the duration of the experiment.Good correlation can be observed between these two values, giving a Pearson correlation coefficient (r) of 0.7. A more detailed view of the morphology is shown by SEM analysis in Fig. \ref{fig:PBS_SEM}, where several types of dissolution behaviour can be seen. \\

\begin{figure}[htb]
	\centering
	\includegraphics[width=0.5\linewidth]{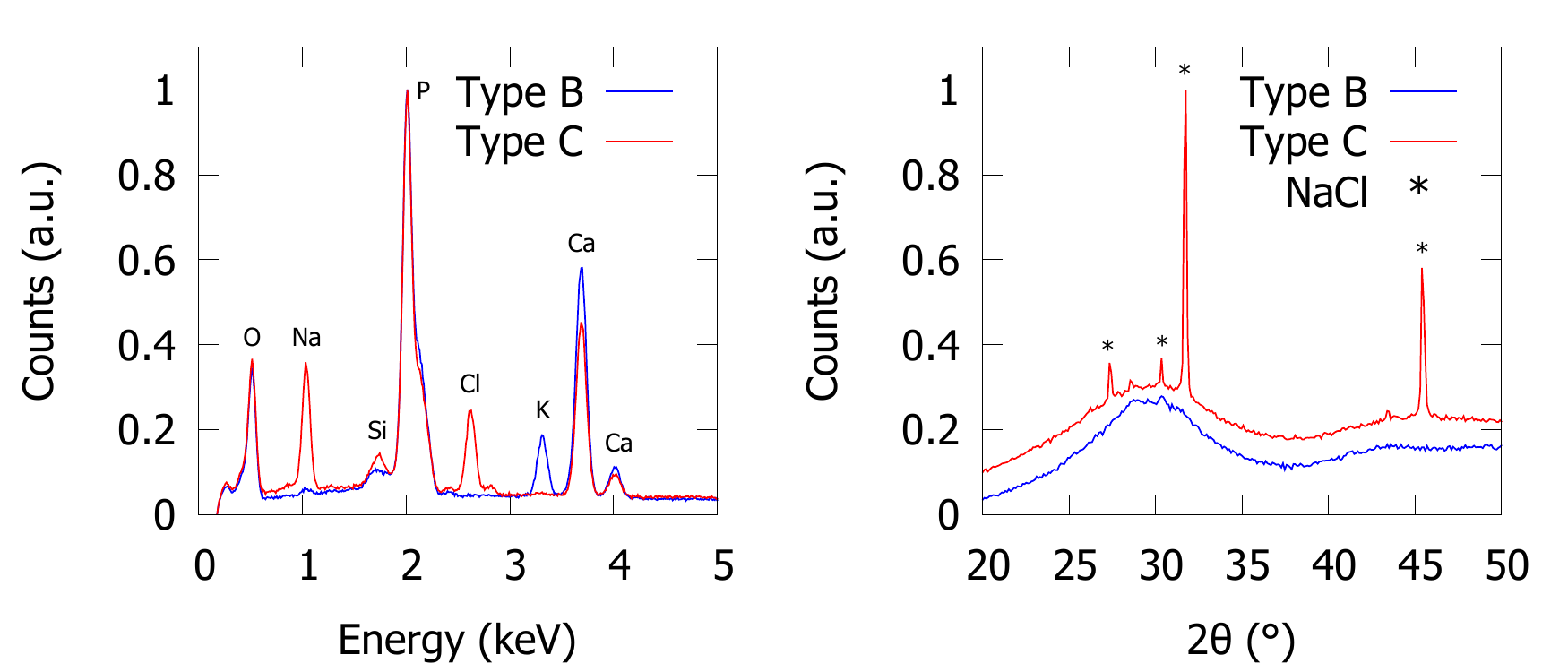}	
	\caption{EDX spectra (left), and XRD patterns (right), for representative samples displaying type B (CaP layer formation) and type C (NaCl layer formation) behaviour. XRD patterns are offset vertically for clarity.}
	\label{fig:EDX_XRD}
\end{figure}

Type A dissolution  (Fig. \ref{fig:PBS_SEM}a) consisted of etch pits in the glass surface, and was seen by EDX to be depleted in Ca compared with the original glass. Type A dissolution was observed at dissolution times below $t_{trans}$, as well as at the end of the study (88 days) for P50Ca40 glass in pH 3 or 5. Type B dissolution (Fig. \ref{fig:PBS_SEM}c-f) consisted of formation of a layer rich in Ca but also K, and depleted in Na. This layer was be identified by EDX but could not be detected by XRD (Fig. \ref{fig:EDX_XRD}), suggesting that the layer had very little long-range crystalline order \cite{Franks2000}. The presence of K in this layer was attributed to the PBS used, the original glasses did not contain any, however the solution was rich in potassium. The cross section shown in Fig. \ref{fig:PBS_SEM} revealed some information about the morphology of this layer, where a dense initial portion was seen, below a more porous section. Fig. \ref{fig:PBS_SEM}f is also interesting to note, as the remnants of type A etch pits (lighter lines across the surface) were visible below the more porous type B layer. Type C dissolution (Fig. \ref{fig:PBS_SEM}b) consisted of formation of a NaCl layer on the glass surface, which could be identified by EDX and XRD (Fig. \ref{fig:EDX_XRD}). This NaCl layer was only observed at dissolution times slightly above $t_{trans}$, but not at the end of the study (88 days).\\

%%%%%%%%%%%%%%%%%%%%%%%%%%%%%%%%%%%%%%%%%%%%%%%%%%%%%%%%%%%%%%%%%%%%%%%%%%%%%%%%%%%%%%%%%%%%%%%%%%%%%%%

\FloatBarrier
\section{Discussion}

\subsection{Structure}

The replacement of Na\textsubscript{2}O by SiO\textsubscript{2} can be attributed to volatilisation of Na from the melt, and replacement by Si from the silica crucible \cite{Shelby2005}. The results observed for characterisation of the phosphate glasses produced (Table \ref{tab:Char}) were consistent with a modified version of the nominal glass model, incorporating the presence of SiO\textsubscript{2} from the crucible: (P\textsubscript{2}O\textsubscript{5})\textsubscript{90-$x$}(CaO)\textsubscript{$x$}(Na\textsubscript{2}O)\textsubscript{7}(SiO\textsubscript{2})\textsubscript{3}. Here the network of phosphate tetrahedra was depolymerised by the addition of other metal ions, which created non-bridging oxygens according to their charge, resulting in the Q\textsuperscript{1}/Q\textsuperscript{2} ratios observed. The increase in density with Ca content was consistent with Ca ions being located at interstitial sites within the phosphate network, forming crosslinks between phosphate glass chains \cite{VanWazer1950, Parsons2006}. The effect of these crosslinks was seen in the increased $T_{g}$ for higher Ca glasses - Ca crosslinks provided greater thermal resistance so increase the $T_{g}$ \cite{Ahmed2004b, Parsons2006}. These crosslinks are also known to provide greater chemical durability, so would be expected to lead to a slower dissolution rate.\\

The presence of small amounts of SiO\textsubscript{2} in phosphate glasses such as the amount observed in this study are known to disrupt the phosphate glass network, which in turn leads to a slight increase in the glass dissolution rate \cite{Patel2006}. This indicates that the presence of this impurity may have increased the glass dissolution rate above that which would be expected without these impurities. With the glasses used here, the SiO\textsubscript{2} content is relatively consistent across the glass compositions at around 3 mol. \%, so this is unlikely to make a significant contribution to the trends observed. Formation of six-fold coordinated Si ions, which can reduce the dissolution rate, has been previously observed in phosphate glasses. However, this requires a higher degree of network connectivity within the phosphate glass network (mostly Q\textsuperscript{3} phosphates) than is present here (only Q\textsuperscript{2} or Q\textsuperscript{1}), as well as a greater concentration of Si, so their presence is not expected in these glass compositions \cite{Tabuchi1994, Ren2018}. Formation of six-fold coordinated Si ions would also result in the appearance of a unique peak at 670 cm\textsuperscript{-1} in the IR spectrum \cite{Tabuchi1994}, which was not observed in our experiments (Fig. \ref{fig:Supp_FTIR}). \\

\subsection{Dissolution kinetics}

\begin{figure}[htb]
	\centering
	\includegraphics[width=0.75\linewidth]{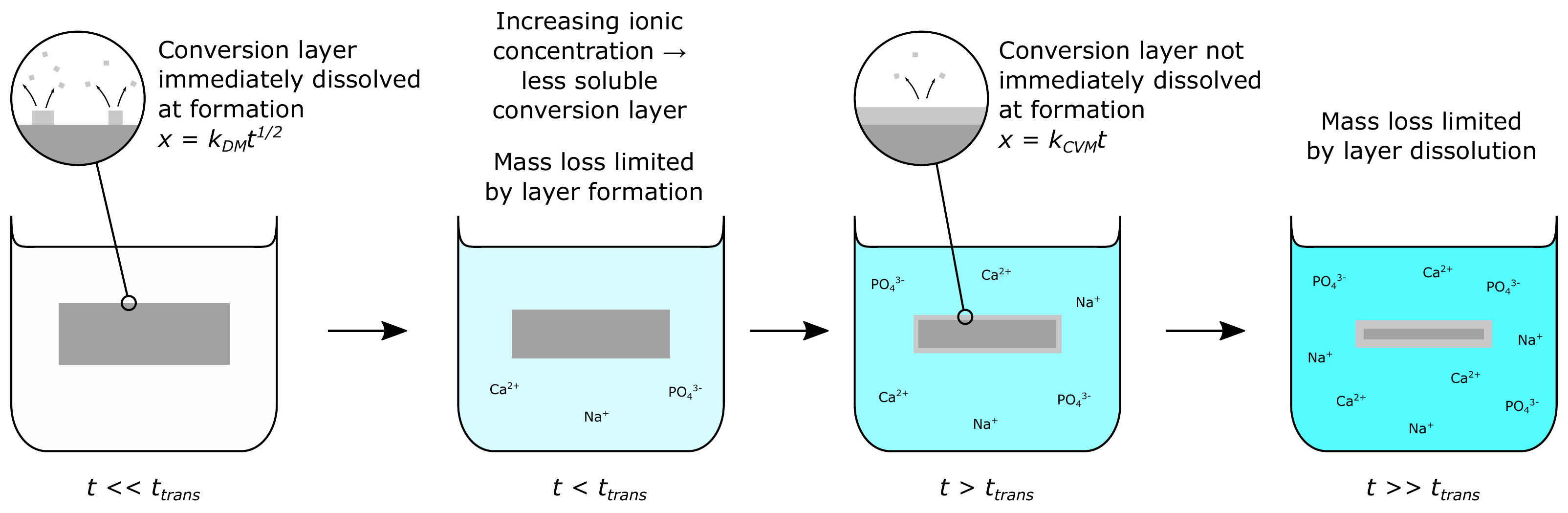}	
	\caption{Schematic diagram illustrating dissolution mechanism, showing diffusion limited dissolution before $t_{trans}$, and surface reaction limited dissolution after $t_{trans}$ when the conversion layer is stable.}
	\label{fig:Mechanism}
\end{figure}

Although phosphate glasses have been researched for several decades now, the mechanisms of their dissolution are still a source of uncertainty \cite{Ma2018}, and it is hoped that the results described here can provide some additional insight into these mechanisms. Here we propose the following explanation for the observed dissolution behaviour, which can be broadly broken down into two main categories of reaction. Firstly we introduce the idea of the formation of a conversion layer, which can include hydration among other reactions. Secondly, layer dissolution reactions encompass dissolution of the various products of conversion layer formation, involving release of ions into solution. It must be noted that both of these reaction categories take place simultaneously throughout the dissolution process, however at various points one or other of these may be rate-limiting. A summary of this mechanism scheme is shown in Fig. \ref{fig:Mechanism}.\\

\textbf{Conversion layer formation} involves reaction of crosslinked phosphate glass anions in the solid state, with reactants that have diffused into the glass such as H\textsubscript{2}O, H\textsuperscript{+} and Cl\textsuperscript{-}. This results in formation of other solid species, usually including the hydration reaction to form hydroxides, but possibly other reactions as well. The hydration reaction (Eq. \ref{eq:Alt-hydration}) has been described by Ma et al. \cite{Ma2018}, defining the hydration of crosslinks between metal cations and two phosphate anions (with $n$ and $m$ phosphate tetrahedra). This is adequate for dissolution in DI water, however the situation is more complex for solutions such as PBS where various other ions are present in solution. Formation of metal hydroxides can be replaced by, or occur in conjunction with, reactions to form solid metal chlorides as described in Eq. \ref{eq:Alt-chlor}. This is particularly significant given the much greater solubility in water of CaCl\textsubscript{2} compared with Ca(OH)\textsubscript{2}.\\

\begin{dmath}
	\label{eq:Alt-hydration}
(P_{n}O_{3n+1})(Ca,Na_{2})_{(n+m+4)/2}(P_{m}O_{3m+1}) + (n\!+\!m\!+\!4)H_{2}O \rightarrow H_{2+n}P_{n}O_{3n+1} + H_{2+m}P_{m}O_{3m+1} + (n\!+\!m\!+\!4)/2(Ca,Na_{2})(OH)_{2}
\end{dmath}

\begin{dmath}
	\label{eq:Alt-chlor}
	(P_{n}O_{3n+1})(Ca,Na_{2})_{(n+m+4)/2}(P_{m}O_{3m+1}) + (n\!+\!m\!+\!4)H^{+} + (n\!+\!m\!+\!4)Cl^{-} \rightarrow H_{2+n}P_{n}O_{3n+1} + H_{2+m}P_{m}O_{3m+1} + (n\!+\!m\!+\!4)/2(Ca,Na_{2})Cl_{2}
\end{dmath}

The reaction front for these conversion layer formation reactions progresses into the glass, following a parabolic time dependence ($t^{\nicefrac{1}{2}}$) due to the diffusive mass transport required to transport reactants to the reaction front. This is supported by the observed formation of a NaCl layer during dissolution, as well as Ca-rich layers (Figs. \ref{fig:PBS_SEM} and \ref{fig:EDX_XRD}).\\

\textbf{Layer dissolution} is the second reaction stage, involving surface dissolution of the solid species formed during the conversion layer formation stage, described in Eqs. \ref{eq:Dis-oh}-\ref{eq:Dis-phos}. As shown in Eq. \ref{eq:Dis-phos}, orthophosphoric acid (n=1) can dissolve directly, while polyphosphates can undergo further hydrolysis. Nevertheless, it is well known that polyphosphates can be released into solution without further hydrolysis \cite{Ma2018, AbouNeel2008a}.\\

\begin{dmath}
	\label{eq:Dis-oh}
	(Ca,Na_{2})(OH)_{2} \rightarrow (Ca,Na_{2})^{2+}	+ 2OH^{-}
\end{dmath}

\begin{dmath}
	\label{eq:Dis-cl}
	(Ca,Na_{2})Cl_{2} \rightarrow (Ca,Na_{2})^{2+}	+ 2Cl^{-}
\end{dmath}

\begin{dmath}
		\label{eq:Dis-phos}
	H_{2+n}P_{n}O_{3n+1} + (n-1)H_{2}O \rightarrow (3n)H^{+}	+ nPO_{4}^{3-}
\end{dmath}

The reaction front for these layer dissolution reactions progresses into the conversion layer, following a linear time dependence ($t$) due to the reaction controlled process. The dissolution of solid species via these reactions will be sensitive to solution conditions, not only pH but various ion concentrations, which can themselves change over time as the glass dissolves. Depending on ion concentrations, reverse reactions can also occur in a highly saturated solution, involving reaction of solution ions to form a precipitate on the glass surface. This is especially important for metal phosphates, as shown in Eq. \ref{eq:Dis-phos-salt}, where metal and phosphate ions (resulting from glass dissolution, or already present in PBS) form metal phosphates. Eq. \ref{eq:Dis-phos-salt} shows formation of simple phosphates, but more complex forms such as octacalcium phosphate or hydroxyapatite are also possible,  as suggested by the morphology of the Ca-rich layer observed (Figs. \ref{fig:PBS_SEM} and \ref{fig:EDX_XRD}) \cite{Franks2000, Barrere2003}.\\

\begin{dmath}
	\label{eq:Dis-phos-salt}
	3(Ca,Na_{2},K_{2})^{2+}	+ 2PO_{4}^{3-} \rightarrow (Ca,Na_{2},K_{2})_{3}(PO_{4})_{2} 
\end{dmath}

These two reaction categories can explain the multi-stage dissolution behaviour observed here and elsewhere. Initially the conversion layer formation reactions, which must occur first to produce soluble salts, are rate-limiting due to fast dissolution of the small amounts of hydrated phosphates and metal hydroxides or chlorides produced. As this reaction is limited by mass transport of reactants to the reaction front, this leads to the initial $t^{\nicefrac{1}{2}}$ dependence of the reaction progress $x$. As the conversion layer dissolves as fast as it is formed, there is no stable conversion layer present on the surface. In later stages ($t > t_{trans}$) the layer dissolution reactions determine the glass dissolution rate, giving rise to the linear $t$ dependence due to assumed linear reaction kinetics. Here the conversion layer is produced faster than it can be dissolved, leading to buildup of the layer on the surface, as illustrated in Fig. \ref{fig:MechSchem}. Despite encompassing dissolution of several layer species (Eqs. \ref{eq:Dis-oh}-\ref{eq:Dis-phos}), as well as deposition reactions (Eq. \ref{eq:Dis-phos-salt}), glass dissolution controlled by dissolution of the conversion layer can be adequately described by Eq. \ref{eq:Mod-kCVM} of the model. Assuming Eqs. \ref{eq:Dis-oh}-\ref{eq:Dis-phos-salt} proceed via linear reaction kinetics, $k_{CVM}$ is simply the sum of equivalent parameters for individual reactions. $k$ for deposition reactions will be negative, and if deposition outweighs mass loss, $k_{CVM}$ will be negative, leading to overall mass gain.\\

The effect of Ca on the dissolution rate in the initial conversion layer formation stage seen here is consistent with multiple previous works, where increased Ca content reduced the dissolution rate \cite{Ahmed2004b, Knowles2003, AbouNeel2009}. Higher Ca glasses were shown to have higher density (Table \ref{tab:Char}), indicating that Ca ions were located at interstitial sites between phosphate glass chains, reducing the free volume \cite{Shelby2005}. In this diffusion controlled stage, this resulted in reduced dissolution (lower $k_{DM}$) by blocking of interstitial diffusion pathways. The greater strength of the Ca-O-P crosslinks compared with P-O-P bonds is another reason for the reduced dissolution rate of higher Ca glasses \cite{Bunker1984}. The rate constant governing this diffusion stage ($k_{DM}$) was shown to have a significant dependence on the starting solution pH (Fig. \ref{fig:Mass_fits}), with large increases in $k_{DM}$ observed as the pH of the solution was reduced. This is consistent with a diffusion mechanism involving H\textsuperscript{+} diffusion into the glass, which would be increased by a larger H\textsuperscript{+} concentration in the solution.\\

The later stages of dissolution are limited by the surface reactions (Eqs. \ref{eq:Dis-oh}-\ref{eq:Dis-phos-salt}), leading to the formation of a stable conversion layer, which persists as it is formed faster than it dissolves. The effect of Ca on the reaction rate here also displays reduced dissolution with increasing Ca content. This can be explained in terms of the hydration reaction (Eq. \ref{eq:Alt-hydration}), which indicates that production of the less soluble Ca(OH)\textsubscript{2} would dominate over NaOH as the Ca/Na ratio increases. Similarly, when PO\textsubscript{4}\textsuperscript{3-} concentration is sufficient, precipitation of phosphates according to Eq. \ref{eq:Dis-phos-salt} would prefer less soluble calcium phosphates, further reducing the mass loss. The effect of pH in this layer dissolution stage is complex, however in general lower pH increases the reaction rate constant $k_{CVM}$, which can be attributed to increased solubility of the conversion layer species in more acidic conditions.\\

The transition time between these two stages is perhaps the least well understood component of the non-linear dissolution of phosphate glasses. Bunker et al. described a rough correlation between glass durability and transition time, also noting that it was also sensitive to pH \cite{Bunker1984}, while Ma et al. found that the transition time only varied for glasses with short phosphate chains (i.e. higher O/P and Q\textsuperscript{1}/Q\textsuperscript{2} ratios), finding a similar correlation to Bunker et al., but also commenting that it may be related to the nature of the surface layer \cite{Ma2017}. In this work we observed a similar correlation to Bunker et al. for dissolution in DI water (more soluble glasses have a shorter transition time), however this trend was absent in PBS and was reversed in pH-adjusted PBS (pH = 3 or 5).\\ 

Under the mechanism proposed here, the transition from the conversion layer formation stage to the layer dissolution stage is controlled predominantly by the nature of the conversion layer (i.e. the combination of solid species formed by Eqs. \ref{eq:Alt-hydration}-\ref{eq:Alt-chlor}) and its solubility in the surrounding solution, which will be heavily dependent on pH and ion concentrations. As the solution chemistry changes during dissolution, increases in dissolved metal (Ca, Na), hydroxide, and phosphate ions will reduce the dissolution rate of the conversion layer. Once the dissolution rate is lower than the formation rate, the conversion layer is stabilised, and the rate limiting reaction will become the layer dissolution, leading to a change from parabolic ($t^{\nicefrac{1}{2}}$) to linear ($t$) reaction kinetics. When this stable layer is dominated by (Ca,Na\textsubscript{2})(OH)\textsubscript{2} or (Ca,Na\textsubscript{2})Cl\textsubscript{2}, or deposited phosphates (Eq. \ref{eq:Dis-phos-salt}), it would be expected to be optically opaque as observed in Fig. \ref{fig:PBS_morph}, explaining the correlation between $t_{trans}$ and the time for observation of an opaque layer (Fig. \ref{fig:t_trans_comp}). The behaviour of P50Ca40 glass seemed to contradict this however, displaying transition to the linear layer dissolution kinetics in DI water, and PBS at pH 3 or 5, without formation of an opaque layer. This can be explained by differences in the phosphate anions present. Based on the glass composition, the number average chain length can be calculated \cite{Bunker1984}; for P50Ca40, P45Ca45, and P40Ca50 the average chain length (n) is predicted to be $\infty$, 9.1 and 4.0 respectively. This means that in acidic conditions, layer dissolution from P50Ca40 may be limited by the hydrolysis reaction of longer phosphate chains (Eq. \ref{eq:Dis-phos}, where n is very large) rather than dissolution of other solid species. Transition to this stage would then not result in a visible opaque layer, as the hydrated phosphate anions (H\textsubscript{2+n}P\textsubscript{n}O\textsubscript{3n+1}) that make up the conversion layer are not forming a new crystalline solid but rather their cation crosslinkers are simply being exchanged for H\textsuperscript{+} ions.\\

%%%%%%%%%%%%%%%%%%%%%%%%%%%%%%%%%%%%%%%%%%%%%%%%%%%%%%%%%%%%%%%%%%%%%%%%%%%%%%%%%%%%%%%%%%%%%%%%%%%%%%%

\FloatBarrier
\section{Conclusions}
These results offer new insight into the mechanism of phosphate glass dissolution in various media, in particular the origin of the parabolic time dependence and transition time between the two stages. Understanding the dissolution behaviour of these glasses is of importance when considering their dissolution in medical applications, and the pH dependence of this is particularly of interest in polymer composite applications, where polymer degradation results in formation of lactic acid and acidification.\\

The dissolution behaviour of a range of P\textsubscript{2}O\textsubscript{5}-CaO-Na\textsubscript{2}O glasses was measured in deionised water, PBS, and pH-adjusted PBS at pH 3 or 5. In accordance with previous studies in DI water, increased Ca was observed to significantly reduce the dissolution rate in all conditions. Two-stage behaviour was seen in DI water, PBS, and pH-adjusted PBS, with an initial parabolic time dependence, followed by later linear time dependent reaction progression. A two-stage model similar to those reported previously was adapted for use with the disc-shaped glass samples here.\\

Based on these results, a new dissolution mechanism was proposed to explain the two-stage dissolution behaviour, that takes into account more complex dissolution media. The initial stage involves diffusion of water or ions into the glass, with $t^{\nicefrac{1}{2}}$ dependence, forming a conversion layer consisting of hydrated phosphate anions, and metal hydroxides or chlorides. Once the solution conditions slow the layer dissolution reactions enough so that the conversion layer is stabilised, the layer dissolution reaction becomes rate-limiting and results in linear $t$ dependence. Under this mechanism, the transition time $t_{trans}$ is sensitive both to the nature of the conversion layer, and the solution conditions.\\

%%%%%%%%%%%%%%%%%%%%%%%%%%%%%%%%%%%%%%%%%%%%%%%%%%%%%%%%%%%%%%%%%%%%%%%%%%%%%%%%%%%%%%%%%%%%%%%%%%%%%%%

\FloatBarrier
\section*{Acknowledgements}
The authors thank Lucideon Ltd. for financial support of the project, and for providing facilities for production of phosphate glass materials. In particular, the assistance of Mr Ian Campbell is greatly appreciated. We would also like to thank Dr Abil Aliev at University College London, for collecting the solid-state NMR spectra. RNO would also like to thank the Woolf Fisher Trust and the Cambridge Trust, for provision of a PhD scholarship, and the Armourers and Brasiers' Company for additional funding. The authors would also like to thank Dr Kyung-Ah Kwon, Mr Wayne Skelton-Hough, Dr Christopher Lovell, Mr Simon Griggs, and Mr Robert Cornell for their technical support and helpful discussions throughout the project. Original data for this paper can be found at \url{https://doi.org/10.17863/CAM.50847}.

%%%%%%%%%%%%%%%%%%%%%%%%%%%%%%%%%%%%%%%%%%%%%%%%%%%%%%%%%%%%%%%%%%%%%%%%%%%%%%%%%%%%%%%%%%%%%%%%%%%%%%%

%\section*{References}
\bibliographystyle{ieeetr}

\bibliography{References}

\FloatBarrier

\clearpage
\newpage

\section*{Supplementary Material: Non-linear dissolution mechanisms of P\textsubscript{2}O\textsubscript{5}-CaO-Na\textsubscript{2}O glasses as a function of pH in various aqueous media}
\begin{flushleft}
	Reece N. Oosterbeek, Kalliope I. Margaronis, Xiang C. Zhang, Serena M. Best, Ruth E. Cameron
	\bigskip
\end{flushleft}

\renewcommand{\thetable}{S\arabic{table}}
\setcounter{table}{0}

\renewcommand{\thefigure}{S\arabic{figure}}
\setcounter{figure}{0}

\renewcommand{\thesection}{S}
\setcounter{figure}{0}

\FloatBarrier
\subsection{Glass characterisation}

\begin{figure}[htb]
	\centering
	\includegraphics[width=0.5\linewidth]{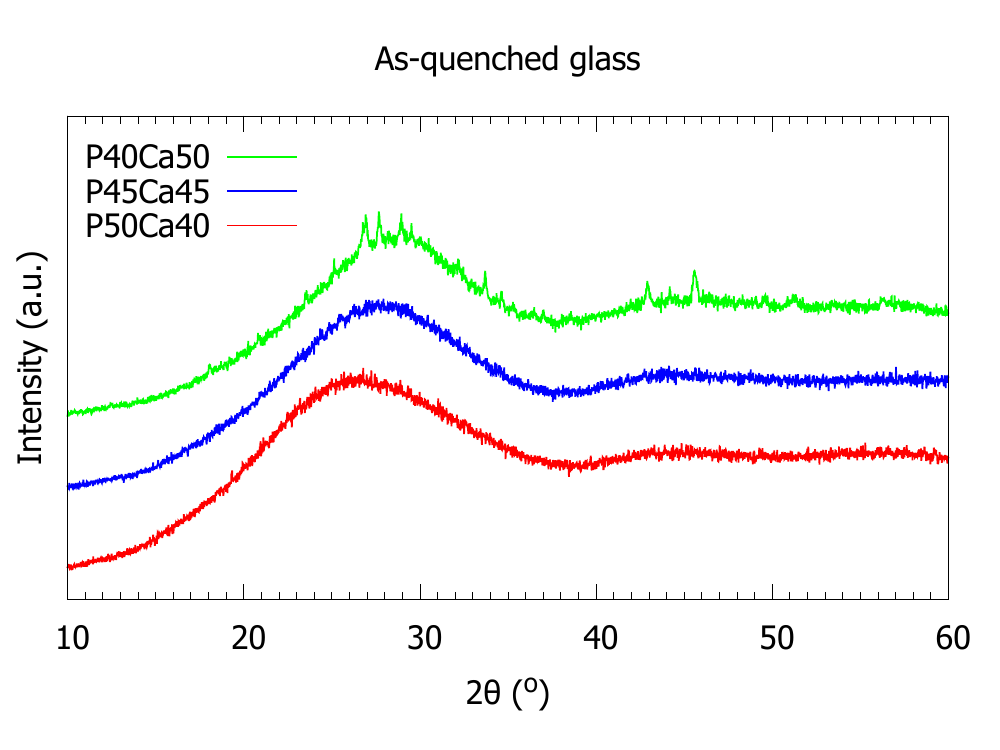}	
	\caption{XRD patterns for as-quenched phosphate glasses. Patterns are offset vertically for clarity.}
	\label{fig:Supp_XRD_Quench}
\end{figure}

\begin{figure}[htb]
	\centering
	\includegraphics[width=0.5\linewidth]{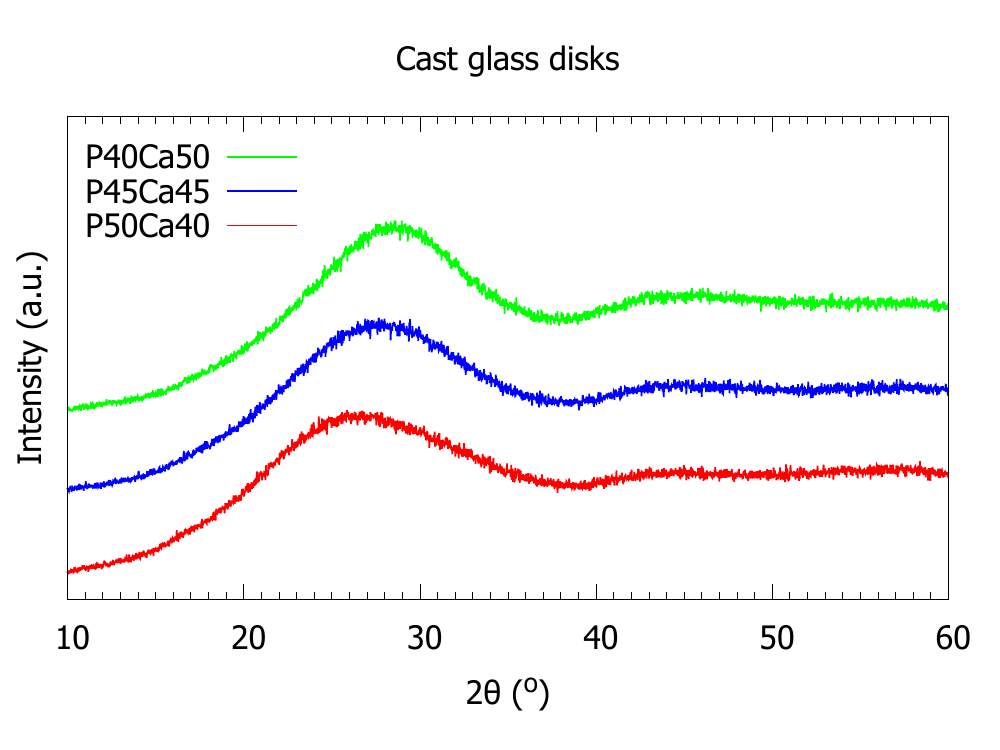}	
	\caption{XRD patterns for cast phosphate glass discs. Patterns are offset vertically for clarity.}
	\label{fig:Supp_XRD_Disk}
\end{figure}

XRD patterns for the phosphate glasses used are shown in Figs. \ref{fig:Supp_XRD_Quench} and \ref{fig:Supp_XRD_Disk}. After initial quenching the P40Ca50 glass showed a small amount of devitrification due to the high amount of modifiers present, as indicated by the small diffraction peaks observed. This was determined to be crystalline $\alpha$-Ca\textsubscript{2}P\textsubscript{2}O\textsubscript{7}. After re-melting and casting this devitrification was not observed, due to the faster cooling during casting preventing crystallisation.

\begin{figure}[htb]
	\centering
	\includegraphics[width=0.5\linewidth]{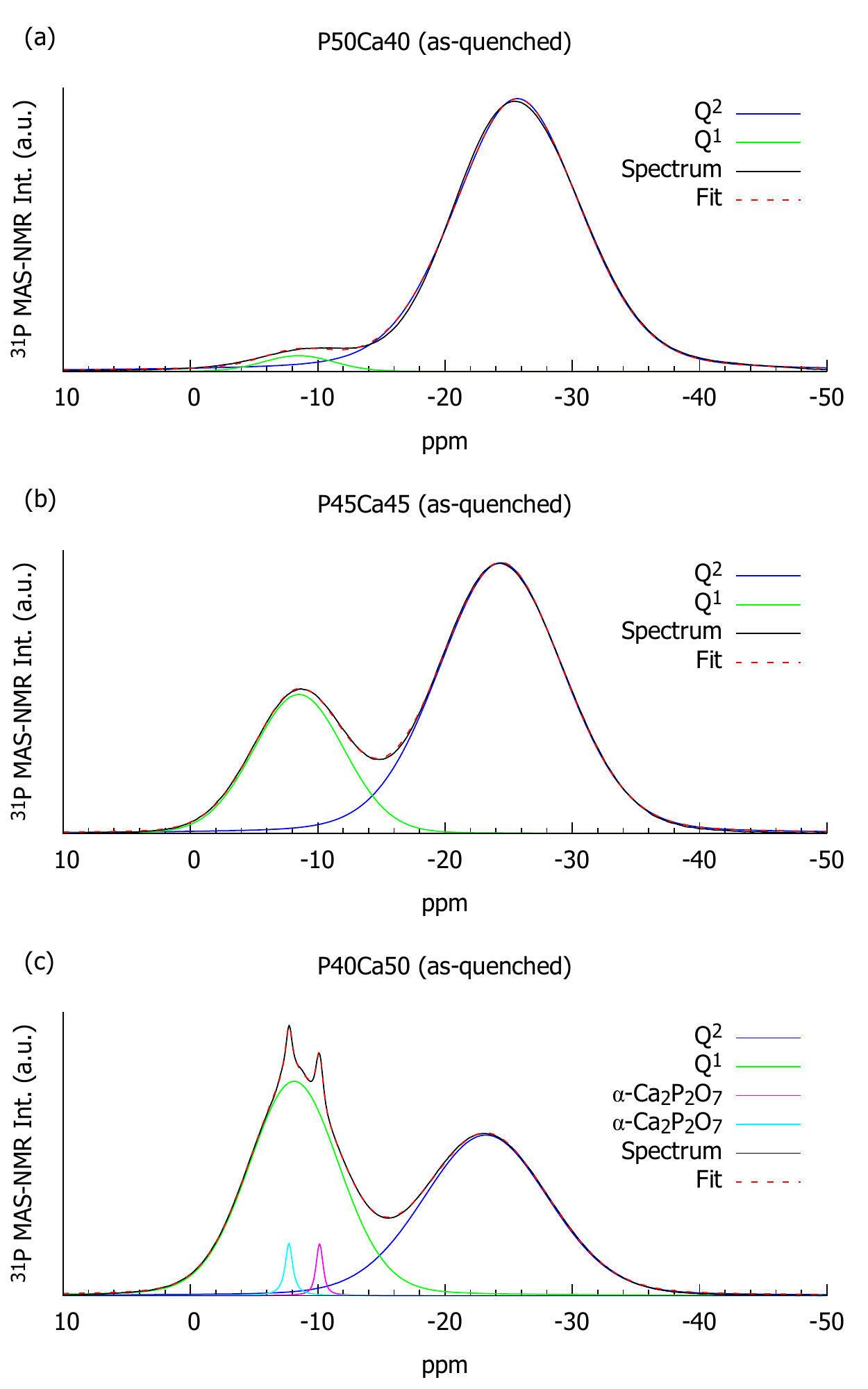}	
	\caption{\textsuperscript{31}P MAS-NMR spectra of as-quenched phosphate glasses, showing measured spectra and fitted peaks.}
	\label{fig:Supp_NMR}
\end{figure}

NMR spectroscopy was carried out on the phosphate glasses in their as-quenched state, before re-melting and casting. These spectra and their deconvolutions are shown in Fig. \ref{fig:Supp_NMR}, with fitted peaks assigned to specific chemical shifts with reference to literature \cite{Brow2000, Fletcher1993, Gras2016}. Q\textsuperscript{1} and Q\textsuperscript{2} phosphate tetrahedra were observed, with the proportion of Q\textsuperscript{1} tetrahedra increasing as more Ca was added to the glass. The high calcium P40Ca50 glass also displayed two small peaks attributed to the crystalline $\alpha$-Ca\textsubscript{2}P\textsubscript{2}O\textsubscript{7}. XRD measurements determined that this devitrified material was not present in the cast disks, so the cast disks can be assumed to consist solely of Q\textsuperscript{1} and Q\textsuperscript{2} phosphate tetrahedra.

\begin{figure}[htb]
	\centering
	\includegraphics[width=0.5\linewidth]{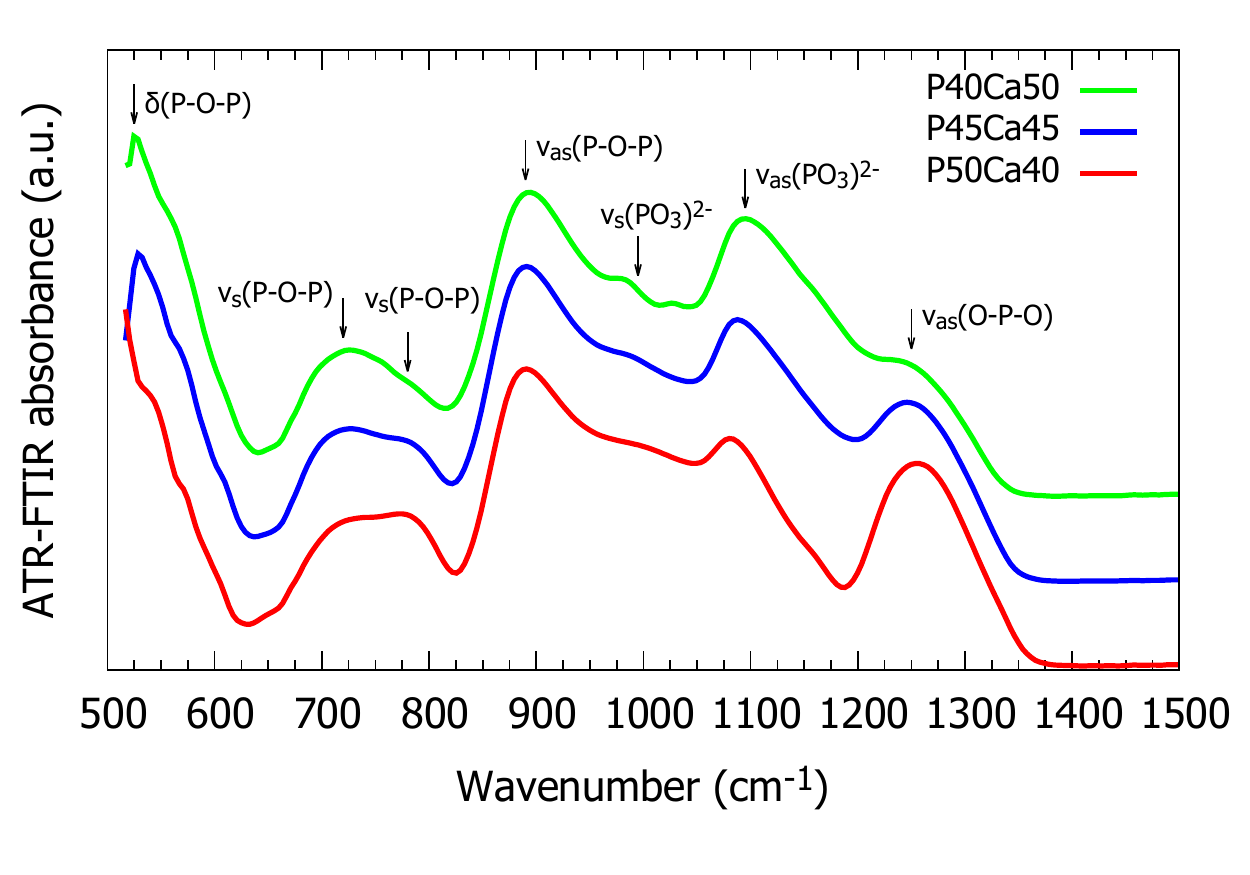}	
	\caption{ATR-FTIR spectra of the phosphate glasses used, with peak assignments shown.}
	\label{fig:Supp_FTIR}
\end{figure}

FTIR (Fourier Transform Infrared Spectroscopy) was carried out on phosphate glass samples using a Bruker Tensor 27 FTIR Spectrometer, in ATR configuration (Attenuated Total Reflectance). These spectra are shown in Fig. \ref{fig:Supp_FTIR}. A background scan was obtained before each sample scan, and three replicates were analysed for each sample. Spectra were collected over the range 520 - 4000 cm\textsuperscript{-1}, with a resolution of 8 cm\textsuperscript{-1}, and 16 scans were collected for each sample.\\ 

\FloatBarrier
\subsection{ISE Calibration}

\begin{figure}[htb]
	\centering
	\includegraphics[width=0.5\linewidth]{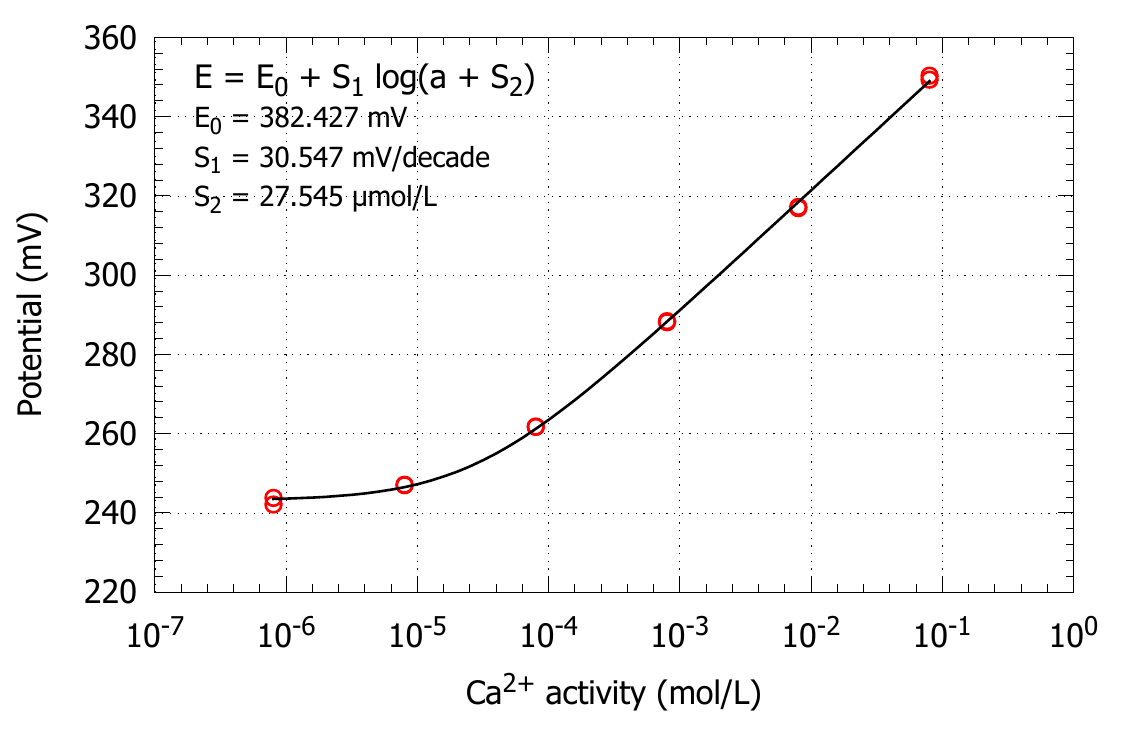}	
	\caption{Typical calibration curve for calcium ISE, using Hanna HI 4004-01 calcium standard.}
	\label{fig:Calibration}
\end{figure}

A typical calibration curve for the calcium ISE is shown in Fig. \ref{fig:Calibration}. Calibration standards were measured twice, the second time in reverse order, to minimise potential drift and hysteresis during measurement. The average of these values was used to fit the calibration curve. A modified Nernst equation was fitted to account for reagent blank determinand as well as interference at low concentration, which occurs with all electrodes \cite{Midgley1977}. Calibration was repeated before measurement of each new timepoint, and the electrode was cleaned if the slope (S\textsubscript{1}) deviated significantly (\textgreater 2 mV) from the theoretical value of 29.6 mV/decade at 25\textcelsius.\\

\FloatBarrier
\subsection{Dissolution model parameters}

\begin{table}[htb]
	\centering
	\caption{\label{tab:glassdiss}{Fitted reaction rate parameters for two-stage models for phosphate glass dissolution.}}
	{\renewcommand{\arraystretch}{1.5}
		\begin{tabularx}{0.84\columnwidth}{ l l l l l l}
			\toprule
			Glass code & Solution & k$_{DM}$ ($\mu$m/day$^{0.5}$) &k$_{CVM}$ ($\mu$m/day) & t$_{trans}$ (days) & R$^{2}$  \\
			\midrule
			P50Ca40 & DI H\textsubscript{2}O & - & 19.4 ($\pm$0.3) & $<$ 0.04 & 0.999 \\
			P45Ca45 & DI H\textsubscript{2}O & 3.5 ($\pm$0.1) & 0.224 ($\pm$0.003) & 4.9 ($\pm$0.5) & 0.989 \\
			P40Ca50 & DI H\textsubscript{2}O & 1.52 ($\pm$0.02) & 0.041 ($\pm$ 0.002) & 13.2 ($\pm$0.7) & 0.982 \\
			P50Ca40 & PBS pH 7 & 6.0 ($\pm$0.3) & 0.78 ($\pm$0.2) & 7 ($\pm$3) & 0.963 \\
			P45Ca45 & PBS pH 7 & 3.9 ($\pm$0.3) & 0.51 ($\pm$0.02) & 7 ($\pm$4) & 0.938 \\
			P40Ca50 & PBS pH 7 & 2.52 ($\pm$0.06) & 0.032 ($\pm$0.005) & 9.3 ($\pm$0.6) & 0.873 \\
			P50Ca40 & PBS pH 5 & 14.9 ($\pm$0.3) & 1.26 ($\pm$0.06) & 22 ($\pm$7) & 0.970 \\
			P45Ca45 & PBS pH 5 & 8.2 ($\pm$0.4) & 1.06 ($\pm$0.02)  & 6 ($\pm$2) & 0.983 \\
			P40Ca50 & PBS pH 5 & 5.1 ($\pm$0.1) & 0.24 ($\pm$0.01) & 10.2 ($\pm0$.8) & 0.959 \\
			P50Ca40 & PBS pH 3 & 34.8 ($\pm$0.8) & 4.0 ($\pm$0.2) & 19.4 ($\pm$0.6) & 0.971 \\
			P45Ca45 & PBS pH 3 & 37.2 ($\pm$0.5) & 1.04 ($\pm$0.08) & 17.2 ($\pm$0.8) & 0.973 \\
			P40Ca50 & PBS pH 3 & 24.0 ($\pm$0.5) & 1.12 ($\pm$0.05) & 10.5 ($\pm$0.8) & 0.964 \\
			\bottomrule
		\end{tabularx}
	}
\end{table}

\vspace{5cm}

\begin{flushleft}

Published journal article:\\
\url{https://doi.org/10.1016/j.jeurceramsoc.2020.08.076}\\
\vspace{0.5cm}
Copyright \textcopyright\ \href{https://creativecommons.org/licenses/by-nc-nd/2.0/uk/}{CC-BY-NC-ND}\\
\vspace{0.2cm}
\includegraphics[width=0.2\linewidth]{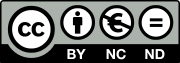}	

\end{flushleft}

\end{document}